\documentclass[12]{article}
\pdfoutput=1
\usepackage{mathrsfs}
\usepackage[numbers,sort&compress]{natbib}
\usepackage{amsmath}
\usepackage{graphicx} 
\usepackage{bmpsize}
\usepackage{enumerate}
\usepackage{url} 

\usepackage[svgnames]{xcolor}
\usepackage{listings}
\begin{document}

\def\spacingset#1{\renewcommand{\baselinestretch}%
{#1}\small\normalsize} \spacingset{1}

\title{Supplemental Studies for Simultaneous Goodness-of-Fit Testing}

\author{Dr. Wolfgang Rolke,
Dept. of Mathematical Sciences, University of Puerto Rico - Mayaguez}

\maketitle

\bigskip
\begin{abstract}
Testing to see whether a given data set comes from some specified distribution is among the oldest 
types of problems in Statistics. Many such tests have been developed and their performance studied. The general result has been that while a certain test might perform well, aka have good power, in one situation it will fail badly in others. This is not a surprise given the great many ways in which a distribution can differ from the one specified in the null hypothesis. It is therefore very difficult to decide a priori which test to use. The obvious solution is not to rely on any one test but to run several of them. This however leads to the problem of simultaneous inference, that is, if several tests are done even if the null hypothesis were true, one of them is likely to reject it anyway just by random chance. In this paper we present a method that yields a p value that is uniform under the null hypothesis no matter how many tests are run. This is achieved by adjusting the p value via simulation. We present a number of simulation studies that show the uniformity of the p value and others that show that this test is superior to any one test if the power is averaged over a large number of cases.
\end{abstract}

\vfill

\newpage
\spacingset{1.5}
\section{Introduction}

A goodness-of-fit test is concerned with the question whether a data set has been generated by a certain distribution. It has a null
hypothesis of the form \(H_0: F=F_0\), where \(F_0\) is a cumulative 
distribution function. For example, one might wish to test whether a data set
comes from a standard normal distribution. An obvious and usually more
useful extension is to test \(H_0: F \in \mathscr{F}_0\) where \(\mathscr{F}_0\) is a family of
distributions but without specifying the parameters. So one might wish
to test whether a data set comes from a normal distribution but without
specifying the mean and standard deviation. 

As described above a goodness-of-fit test is a hypothesis test in the
Fisherian sense of testing whether the data is in agreement with a
model. The main issue with this approach is that it does not allow one to
decide which of two tests is better, that is has the higher power. To solve this problem
Neyman and Pearson in the 1930s introduced the concept of an alternative
hypothesis, and most tests done today follow more closely the
Neyman-Pearson description, although they often are a hybrid of both.
The original Fisherian test survives mostly in the goodness-of-fit
problem, because here the obvious alternative is \(H_a: F \not\in \mathscr{F}_0\), a
space so huge as to be useless for power calculations. 

The goodness-of-fit (gof) test is one of the oldest and most studied problems in
Statistics. For an introduction to Statistics and hypothesis testing in
general see \cite{casella2002} or \cite{bickel2015}. For discussions of the many
goodness-of-fit tests available see \cite{agostini1986}, \cite{raynor2009}, \cite{zhang2002} and  \cite{thas2010}. \cite{thas2010} has an extensive list of references on the subject. 

\section{The Tests}

Many goodness-of-fit methods have been developed over time. In this section we will briefly discuss those currently implemented in our method. Let $n$ be the sample size and $x_1,..,x_n$ the ordered data set. Let $F$ be the distribution function specified under the null hypothesis, either with all parameters fixed or with parameters estimated from the data.

\begin{enumerate}

  \item \textbf{Chi-square tests:} this is the oldest gof test, dating back to Pearson \cite{pearson1900}. It was originally invented for discrete data and applying it to continuous data requires that the data be binned. This can be done in an infinite number of ways, and many studies have investigated the effect of the binning on both the null distribution and on the power of this test, see for example \cite{watson1958},
	\cite{berkson1980}, \cite{bogdan1995}, \cite{dahiya1973}, \cite{greenwood1996}, \cite{harrison1985}, \cite{kallenberg1985}, \cite{koehler1990}, \cite{ooosterhoff1985}, \cite{mineo1979}, \cite{quine1985} and \cite{voinov2013}. 
	
	The two most commonly used methods are bins of equal size (except for a few bins with exceptionally low numbers of observations) and bins with equal counts or probability under the null. Also numerous formulas for the number of bins have been developed. \cite{rolke2020} discuss a novel binning scheme and show that relatively few bins are often best. We will use their binning scheme with $k=5+m$ bins and $\kappa=0.5$, where m is the number of parameters estimated from the data. The routine can also use equal size and/or equal probability bins with a number of bins chosen by the user.
	
	A large number of tests are based on a measure of the distance between the distribution function specified under the null hypothesis and the empirical distribution function: 
	
  \item \textbf{Kolmogorov-Smirnov KS:} next to the chi-square test this is clearly the most commonly employed gof test. The test statistic is give by
	
	$$KS=\max\{i/n-F(x_i),F(x_i)-(i-1)/n \}$$
	
	For further discussions see \cite{birnbaum1952}, \cite{goodman1954} and \cite{massey1951}.
	
	\item \textbf{Anderson-Darling AD} 
		
		$$AD = -n-\frac1n \sum_{i=1}^n (2i-1)\left(\log(F(x_i)+\log(1-F(x_{n+1-i}) \right)$$
		
		For further details see \cite{anderson1952} and \cite{anderson1954}
		
	\item \textbf{Cramer-vonMises CM} 
		
		$$CM = \frac1{12n}+ \sum_{i=1}^n \left(\frac{2i-1}{2n}-F(x_i)\right)^2$$
		
		For further details see \cite{anderson1962}
		
	\item \textbf{Wilson W} 
		
		$$CM = \frac1{12n}+ \sum_{i=1}^n \left(\frac{2i-1}{2n}-F(x_i)\right)^2 - n (\bar{F(x}-\frac12)^2$$

	\cite{zhang2002} studies three test statistics based on likelihood ratios:	
			
	\item \textbf{Zhang ZK}
		
		$$ZK = \max\{(i-0.5)\log \frac{i-0.5}{nF(x_i)}+(n-i+0.5)\log \frac{n-i+0.5}{n(1-F(x_i))}  \}$$
		
	\item \textbf{Zhang ZA}
	
		$$ZA = (-1) \sum_{i=1}^n \frac{\log F(x_i)}{n-i+0.5} +\frac{\log 1-F(x_i)}{i-0.5}$$				
		
		\item \textbf{Zhang ZC}
		
		$$ZC = \sum_{i=1}^n \left(\frac{\log (1/F(x_i)-1)}{(n-0.5)(i-0.75)-1}\right)^2$$
		
Next we have a method based on the correlation between the ordered data and and the quantiles of the distribution:
		
Let $p_i,i=1,..,n$ be the points calculated by the R routine \textit{ppoints}, see \cite{blom1958}, and let $q_i=F^{-1}(p_i)$, then
		
		\item \textbf{Probability Plot Correlation Coefficient ppcc}
		
		$$pp = 1-cor(x,q)$$
		
This test was discussed in \cite{filliben1975}. Note we changed to 1-cor(x,q) from the usual definition cor(x,q) so that large values of the test statistic will lead to rejection of the null hypothesis. The same is also true for the next test:
		
		\item \textbf{Shapiro-Wilk  SW}
		
This test was specifically developed for the normal distribution, see \cite{shapiro1965}. We use the R routine \emph{shapiro.test} to find the value of the test statistic.

Also designed specifically for the normal distribution is the

  \item \textbf{Jarque-Bera test JB}
			
$$
\begin{aligned}
&\hat{\mu}_k = \frac1n \sum_{i=1}^n (x_i-\bar{x})^k \\
&S = \frac{\hat{\mu}_3}{\hat{\mu}_2^{3/2}}  \\
&K = \frac{\hat{\mu}_4}{\hat{\mu}_2^2}  \\
&JB  =\frac{n}6(S^2+(K-3)^2/4) \\
\end{aligned}
$$			

see \cite{jarque1980}

\item \textbf{Neyman Smooth tests}
		
Finally we use the Neyman smooth tests implemented in the R package \emph{ddst} for the cases of the normal, uniform and exponential distributions, see \cite{neyman1937}, \cite{ledwina1994} and \cite{inglot2006}. 

\end{enumerate}
	
Some of these tests such as KS and AD are distribution-free when the null hypothesis specifies the distribution completely. That is, in those cases the test statistics have distributions that are known analytically. In our routine, however, we will not make use of this feature. This is because for others the null distribution always has to be found via simulation, and simply doing so for all tests adds a negligible computational effort. Moreover, even those tests loose their distribution-free property in the more interesting case where only the form of the distribution is specified but parameters have to be estimated in some way. 

It is very simple to add other tests to our routine. All that is needed is to add some R code to the TS function that calculates the test statistic. However, it is assumed that it is a large value of the test statistic that leads to rejection of the null hypothesis.

\section{p value adjustment}

Let's say we carry out $k$ hypothesis tests $H^1, .., H^k$ and let's assume that all $k$ null hypotheses are in fact true. Let's denote by $P_i$ the p value of the $i^{th}$ test. Before carrying out the test $P_i$ is a random variable, and if the underlying distribution is continuous $P_i \sim U[0,1]$.  We are interested in whether any of the tests rejects the null hypotheses, and we denote by $P$ the p value of this test. Its distribution is given by

$$F_P(p)=Prob(P<p)  =1-Prob(P>p) = 1-Prob(P_i>p;i=1,..,k)$$

If all these tests were independent we would find

$$F_P(p)=1-\prod_{i=1}^k Prob(P_i>p) =  1-(1-p)^k$$

Finally using the probability integral transform we could adjust the p value so that the new p value $F(P)=1-(1-P)^k$ would again have a uniform [0,1] distribution. This is of course the basis for the Bonferroni correction, where one often also uses the Taylor approximation $1-(1-x)^k\approx kx$ and then adjusts the type I error probability of the individual tests to $\alpha/k$.

Clearly though in our case the tests are not independent because it is the same data set in all of them. Therefore we do not know what $F_P$ is. We can however estimate it via simulation: generate a data set according to the distribution under the null hypothesis, possibly using the parameter estimates from the data. Apply each test to the simulated data and find the respective values of the test statistic. Repeat many (say $B=25000$) times. Next generate another data set and find the p values as the percentage of test statistics that are larger than this one for each test. Find the smallest of the p values. Again repeat this $B$ times. Use the empirical distribution function of these simulated p values $\hat{F}_P$ as an estimate of $F_P$. In our routine we also use linear interpolation between the jump points of the empirical distribution function. Now find the p value for the actual data, say $p_D$, and adjust it by calculating $\hat{F}_P(p_D)$.

\begin{figure}
\begin{center}
\includegraphics[width=4.5in]{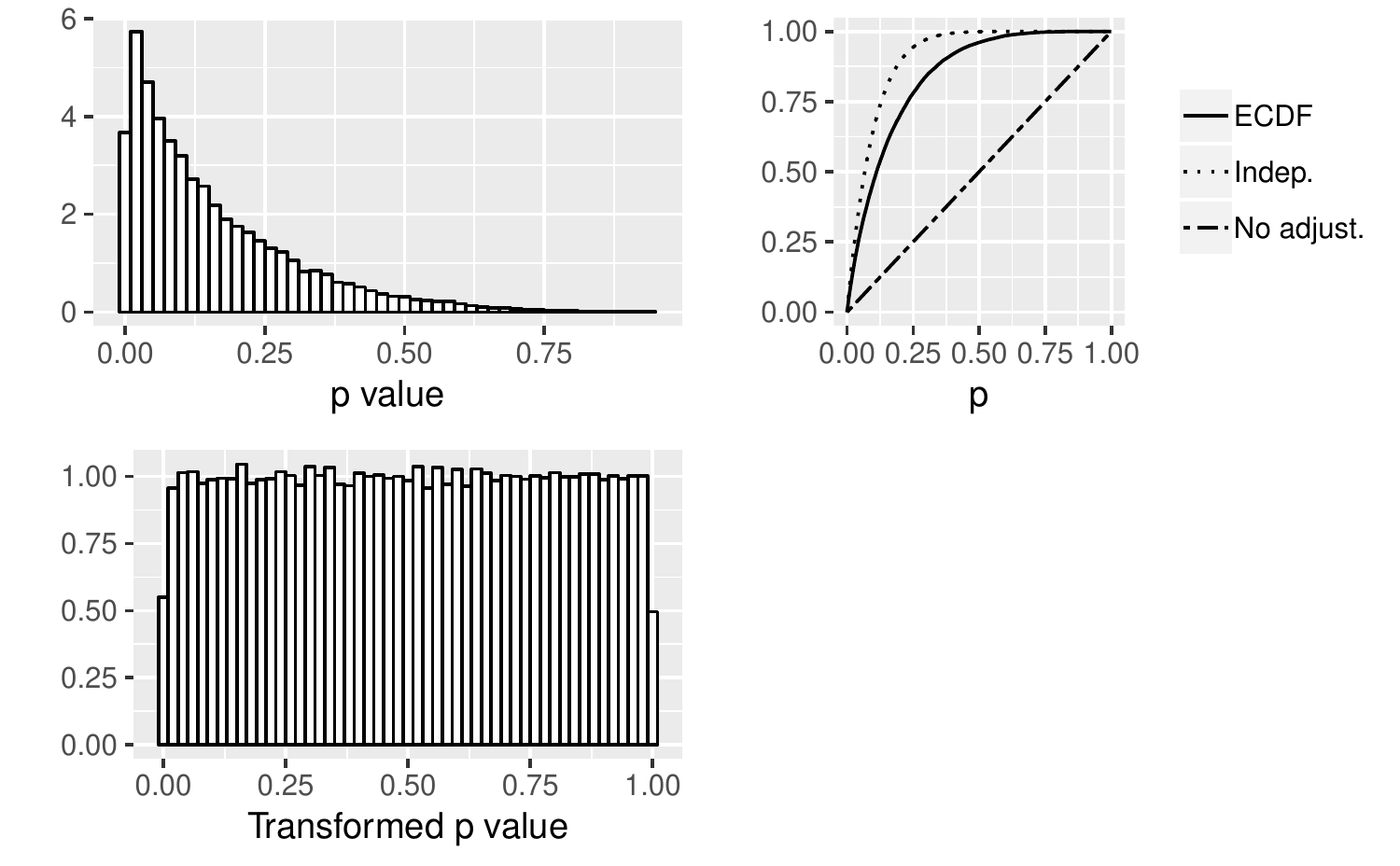}
\end{center}
\caption{Illustration of p-value transformation for multiple comparisons in ANOVA.}
\label{fig:ANOVA}
\end{figure}

This method for adjusting a p value is in fact quite general. As an example, consider the classic problem of pairwise comparisons of group means. As an illustration we generate $100$ observations from a standard normal distribution. Each is assigned at random to one of $5$ groups, so the population group means are in fact all equal to $0$. We then carry out each of the $10$ pairwise comparisons using the two-sample t test and record the smallest p value. The left upper panel of Figure~\ref{fig:ANOVA} shows the histogram of p values, which are clearly far from uniform. Next we find the empirical distribution function, shown in the upper right panel. Here we also add the curves for $y=x$ and $y=1-(1-x)^{10}$, respectively. These correspond to the distribution functions of a uniform [0,1] (no adjustment needed) and the case of independent tests (Bonferroni adjustment). Clearly the pairwise comparison case is somewhat intermediate. Finally the (interpolated) empirical distribution function is applied to the p values, and the lower left panel shows the histogram of transformed p values, now clearly uniform.

This method for adjusting the p value is of course not new, however it has only recently with the availability of fast computers become doable in many problems. For a general discussion of this idea see \cite{buja2006}. For an application to simultaneous confidence bands in quantile-quantile plots see \cite{aldor2013} and an application to goodness-of-fit testing is discussed in \cite{gontscharuk2016}.

\section{Other circumstances}

Our routine also allows for two situations sometimes encountered in practice:

\subsection{Already binned data}

In some fields it is common that the data, although coming from a
continuous distribution, is already binned. This is typically the case,
for example, in high energy physics experiments because of finite
detector resolution. Our routine attempts to `recreate` the original data by spreading it out within the bins. This is done according to the quantile function if one is provided or uniformly if not. 

\subsection{Random sample size}

Another feature often encountered is that the sample size itself is
random. This is the case, for example, if the determining factor was the
time over which an experiment was run. Our routine allows this if the
sample size is a variate from a Poisson distribution with known rate
\(\lambda\), as is often the case.

\section{Performance}

We have carried out a large number of simulation studies to investigate the performance of this method.

\subsection{Type I error}

Because we use simulation to find the distributions of the test statistics under the null hypothesis as well as the distribution of the minimum p value and the p value adjustment, the method will achieve the nominal type I error probability essentially by construction. Nevertheless, table 1 shows the actual type I error probabilities at the nominal 1\%, 5\% and 10\% levels for a number of null hypotheses. Here each simulation is based on $25000$ runs.

\begin{table}
\caption{Actual type I error probabilities for a number of null distributions, sample sizes and nominal type I error probabilities.}
\begin{center}
\begin{tabular}{l|l|r|r|r|r}
\hline
Distribution & Parameters & Sample Size & $1\%$ & $5\%$ & $10\%$\\
\hline
Normal & Fixed & 100 & 1.1 & 5.4 & 10.1\\
\hline
Normal & Fixed & 500 & 1.0 & 5.4 & 10.2\\
\hline
Normal & Fixed & 1000 & 1.0 & 5.0 & 9.8\\
\hline
Normal & Estimated & 100 & 0.9 & 4.9 & 10.0\\
\hline
Normal & Estimated & 500 & 1.2 & 5.2 & 10.1\\
\hline
Normal & Estimated & 1000 & 1.2 & 5.5 & 10.3\\
\hline
Uniform & Fixed & 100 & 1.1 & 4.9 & 10.0\\
\hline
Uniform & Fixed & 500 & 1.1 & 4.7 & 9.7\\
\hline
Uniform & Fixed & 1000 & 0.9 & 4.8 & 10.0\\
\hline
Exponential & Fixed & 100 & 1.2 & 5.2 & 10.0\\
\hline
Exponential & Fixed & 500 & 1.0 & 4.9 & 10.1\\
\hline
Exponential & Fixed & 1000 & 1.1 & 5.2 & 10.3\\
\hline
Exponential & Estimated & 100 & 1.0 & 5.0 & 10.0\\
\hline
Exponential & Estimated & 500 & 1.0 & 4.4 & 9.4\\
\hline
Exponential & Estimated & 1000 & 1.1 & 5.3 & 10.5\\
\hline
Beta & Fixed & 100 & 1.1 & 5.1 & 10.5\\
\hline
Beta & Fixed & 500 & 1.1 & 5.1 & 9.8\\
\hline
Beta & Fixed & 1000 & 1.1 & 4.9 & 9.7\\
\hline
Gamma & Fixed & 100 & 1.0 & 5.1 & 10.2\\
\hline
Gamma & Fixed & 500 & 1.0 & 4.9 & 9.6\\
\hline
Gamma & Fixed & 1000 & 1.1 & 4.9 & 9.9\\
\hline
\end{tabular}
\end{center}
\end{table}

\clearpage
\newpage

\subsection{Power}

Next we discuss a number of case studies for the power of this method. In all of them the sample size is $1000$, the null distribution is found based on $25000$ simulation runs and the power based on $10000$ runs.

The first five cases all specify a normal distribution under the null hypothesis:

\subsubsection{Normal vs t}

\begin{figure}
\begin{center}
\includegraphics[width=4in]{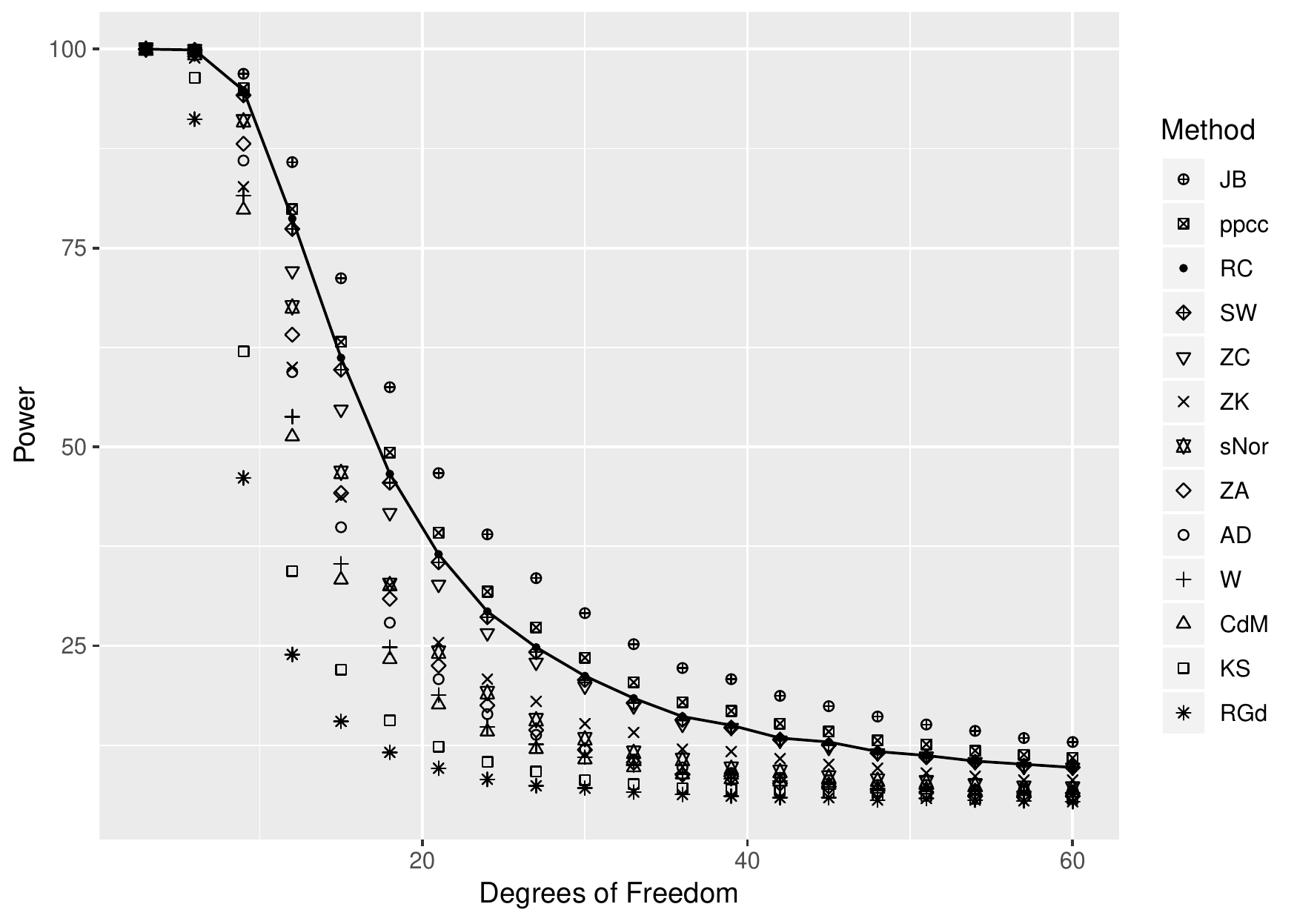}
\end{center}
\caption{Power of various tests if null hypothesis specifies a normal distribution with parameters estimated and the true distribution is a t with degrees of freedom going from 3 to 60.}
\label{fig:Case0}
\end{figure}

Here the mean and standard deviation are estimated via maximum likelihood. The true distribution is a t distribution with $n=3, 6, .. , 60$ degrees of freedom. The power curves are shown in Figure~\ref{fig:Case0}. The method with the highest mean power is JB, followed by ppcc, RC, SW, ZC, ZK, sNor, ZA, AD, W, CdM, KS and RGd. The new RC method is in third place. In all the power graphs it is highlighted by connecting its dots and the legend shows the methods in the order of their mean power. 

\clearpage
\newpage

\begin{figure}
\begin{center}
\includegraphics[width=4.5in]{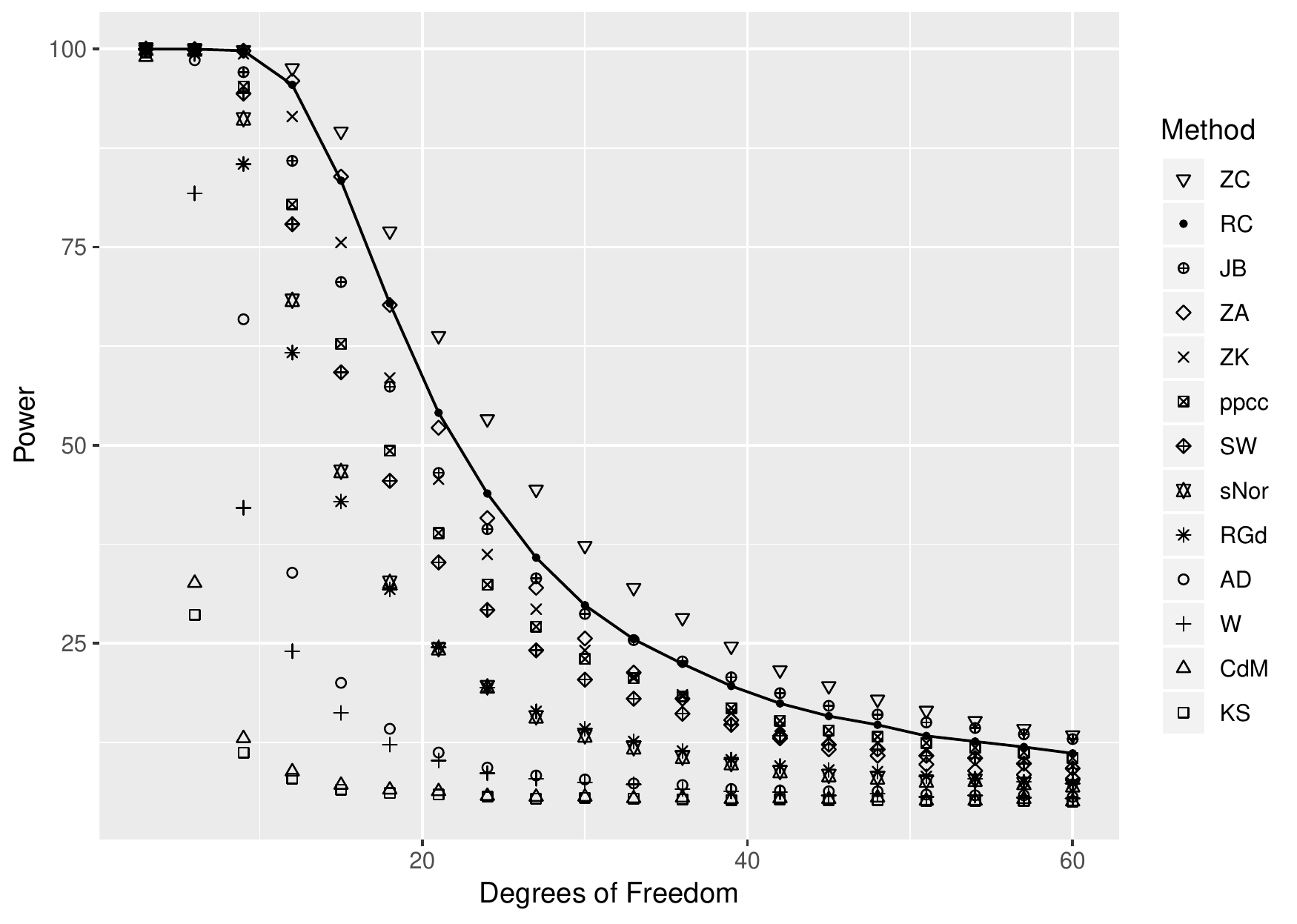}
\end{center}
\caption{Power of various tests if null hypothesis specifies the standard normal distribution and the true distribution is a t with degrees of freedom going from 3 to 60.}
\label{fig:Case1}
\end{figure}

\subsubsection{Normal(0, 1) vs t, Figure~\ref{fig:Case1}}

The setup is the same as above but the now the mean and standard deviation are fixed.  

\clearpage
\newpage

\begin{figure}
\begin{center}
\includegraphics[width=4.5in]{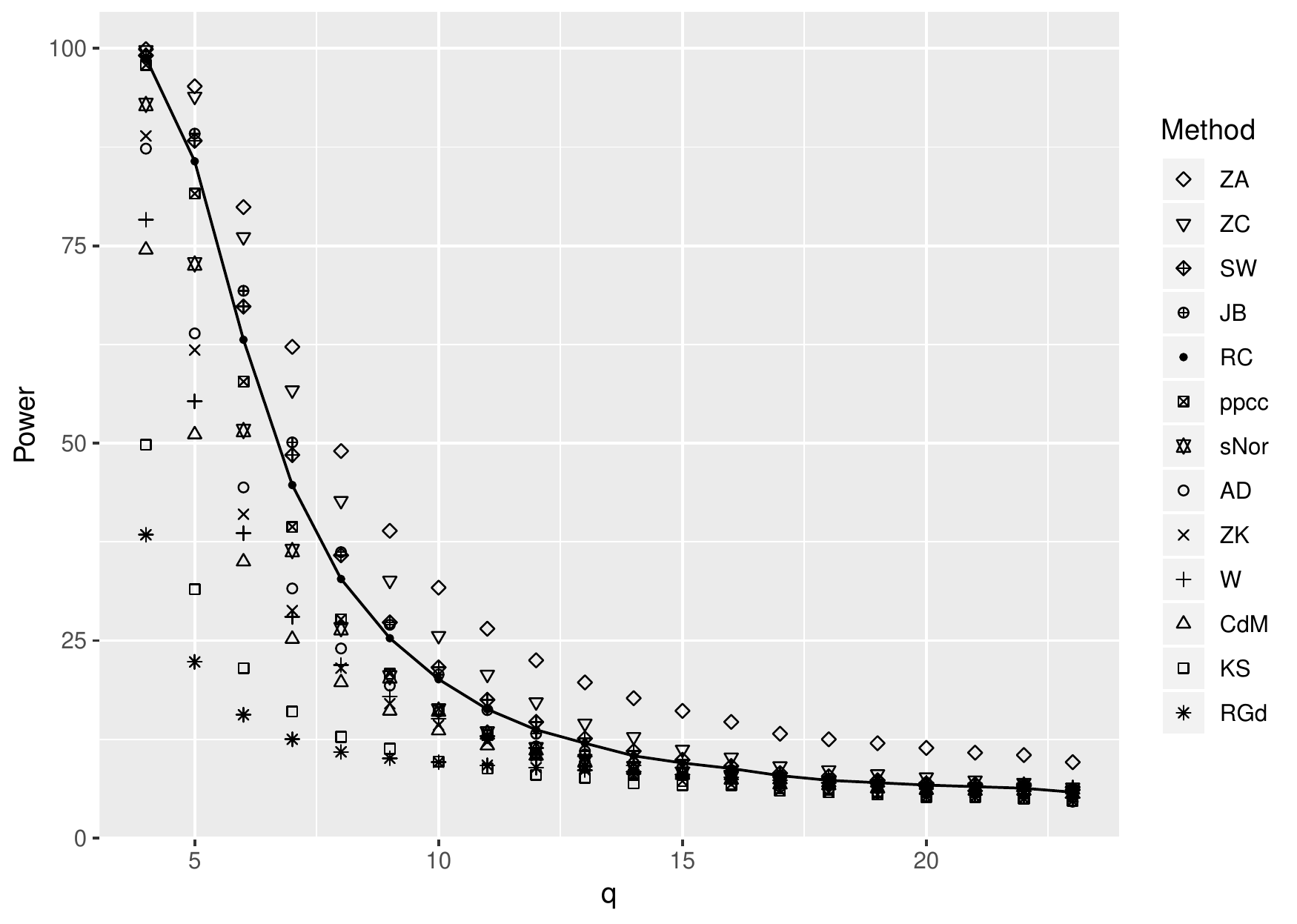}
\end{center}
\caption{Power of various tests if null hypothesis specifies a normal distribution with parameters estimated and the true distribution is a Beta(q,q).}
\label{fig:Case2}
\end{figure}

\subsubsection{Normal vs Beta(q, q), Figure~\ref{fig:Case2}}

Mean and standard deviation are estimated via maximum likelihood. The true distribution is a Beta(q, q) distribution with $q=5:24$. 

\clearpage
\newpage

\begin{figure}
\begin{center}
\includegraphics[width=4.5in]{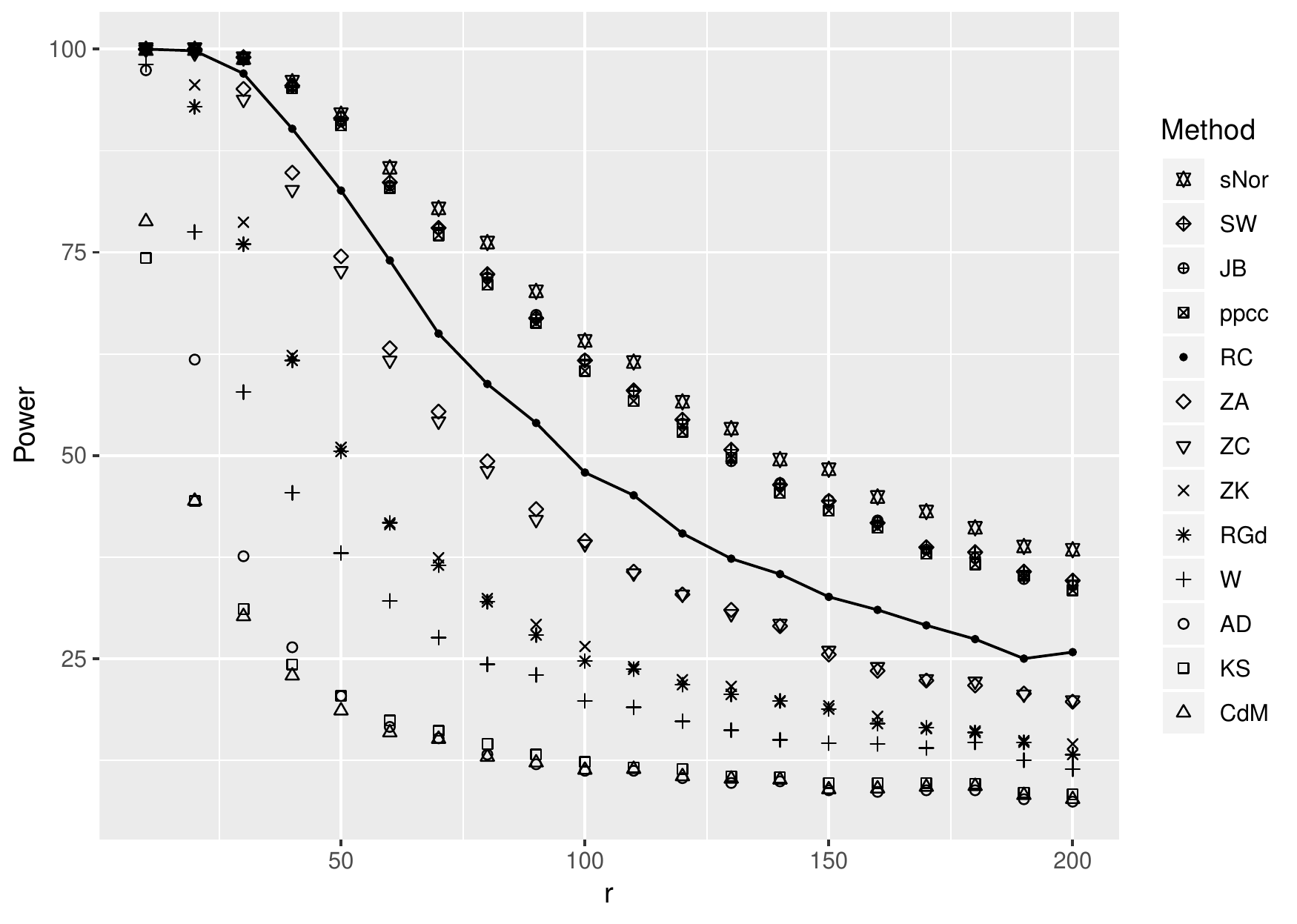}
\end{center}
\caption{Power of various tests if null hypothesis specifies normal distribution with mean and standard deviation r and the true distribution is a Gamma(r, 1).}
\label{fig:Case3}
\end{figure}

\subsubsection{Normal(r, r) vs Gamma(r, 1), Figure~\ref{fig:Case3}}

The mean and variance of the normal distribution under the null are fixed at r and r. The data comes from a Gamma(r ,1) distribution, where $r=2*1:20+3$. The power graph is in Figure~\ref{fig:Case3}. 

\clearpage
\newpage

\begin{figure}
\begin{center}
\includegraphics[width=4.5in]{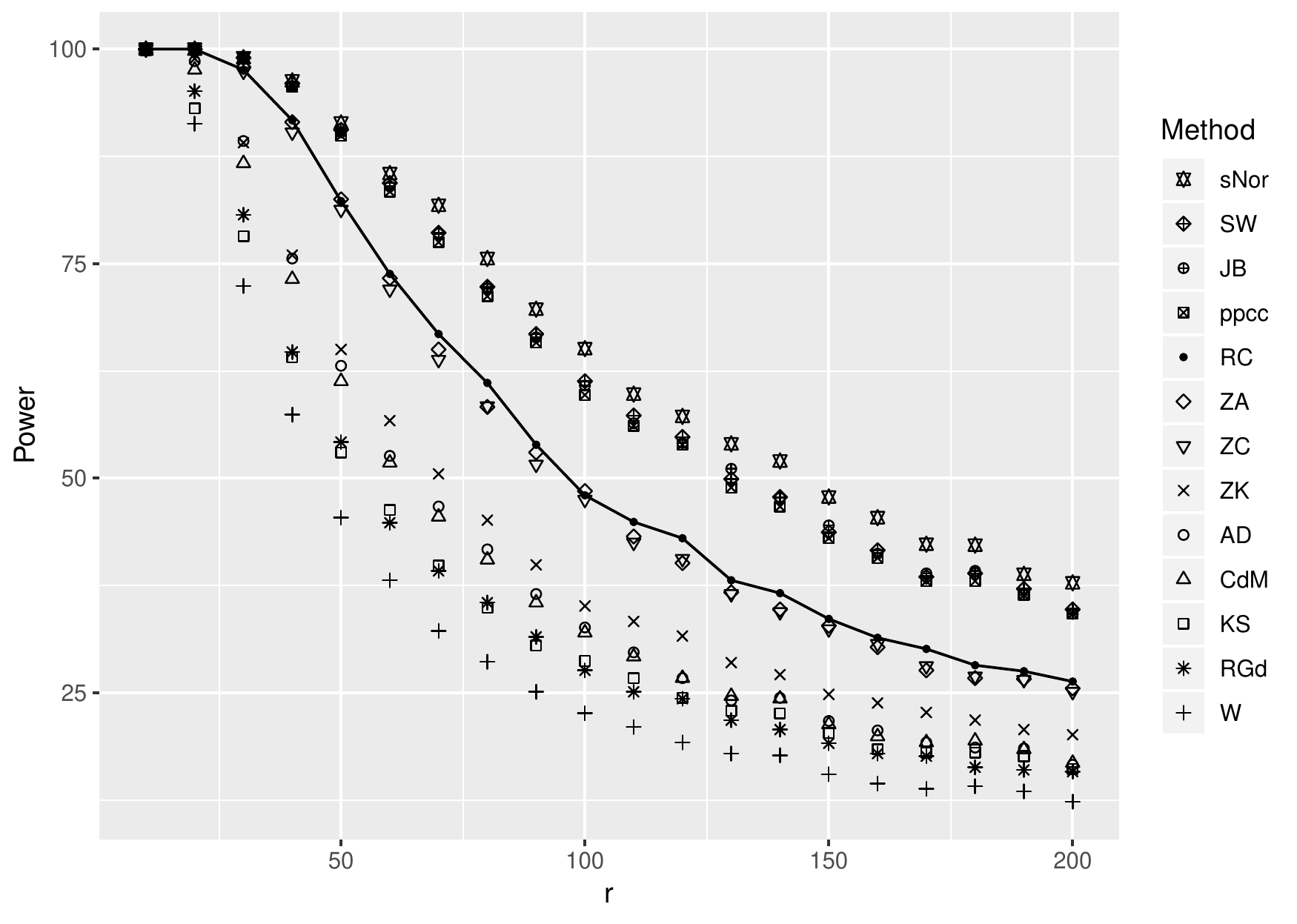}
\end{center}
\caption{Power of various tests if null hypothesis specifies normal distribution with mean and standard deviation estimated and the true distribution is a Gamma(r, 1).}
\label{fig:Case4}
\end{figure}

\subsubsection{Normal vs Gamma(r, 1), Figure~\ref{fig:Case4}}

Same setup as in the previous example, but now the mean and the standard deviation are estimated via maximum likelihood.

\clearpage
\newpage

\begin{figure}
\begin{center}
\includegraphics[width=4.5in]{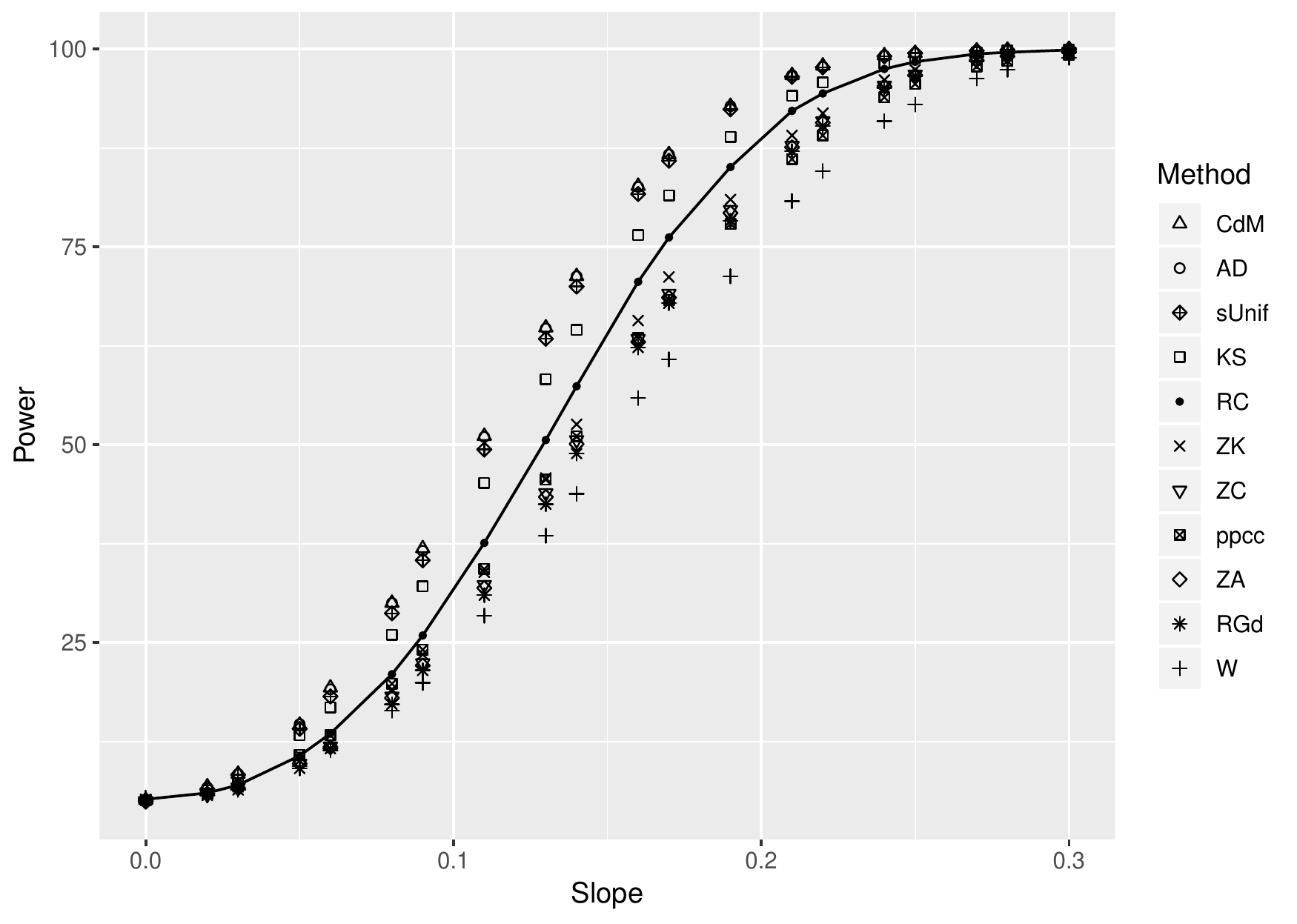}
\end{center}
\caption{Power of various tests if null hypothesis specifies a Uniform [0,1] and the true distribution is linear.}
\label{fig:Case5}
\end{figure}

Next we consider four cases with a uniform distribution under the null hypothesis:

\subsubsection{U[0,1] vs Linear(s), Figure~\ref{fig:Case5}}

The linear density is parametrized as $f(x;s)=2sx+1-s,0<x<1$, so the case $s=0$ corresponds to the U[0,1] distribution. 

\clearpage
\newpage

\begin{figure}
\begin{center}
\includegraphics[width=4.5in]{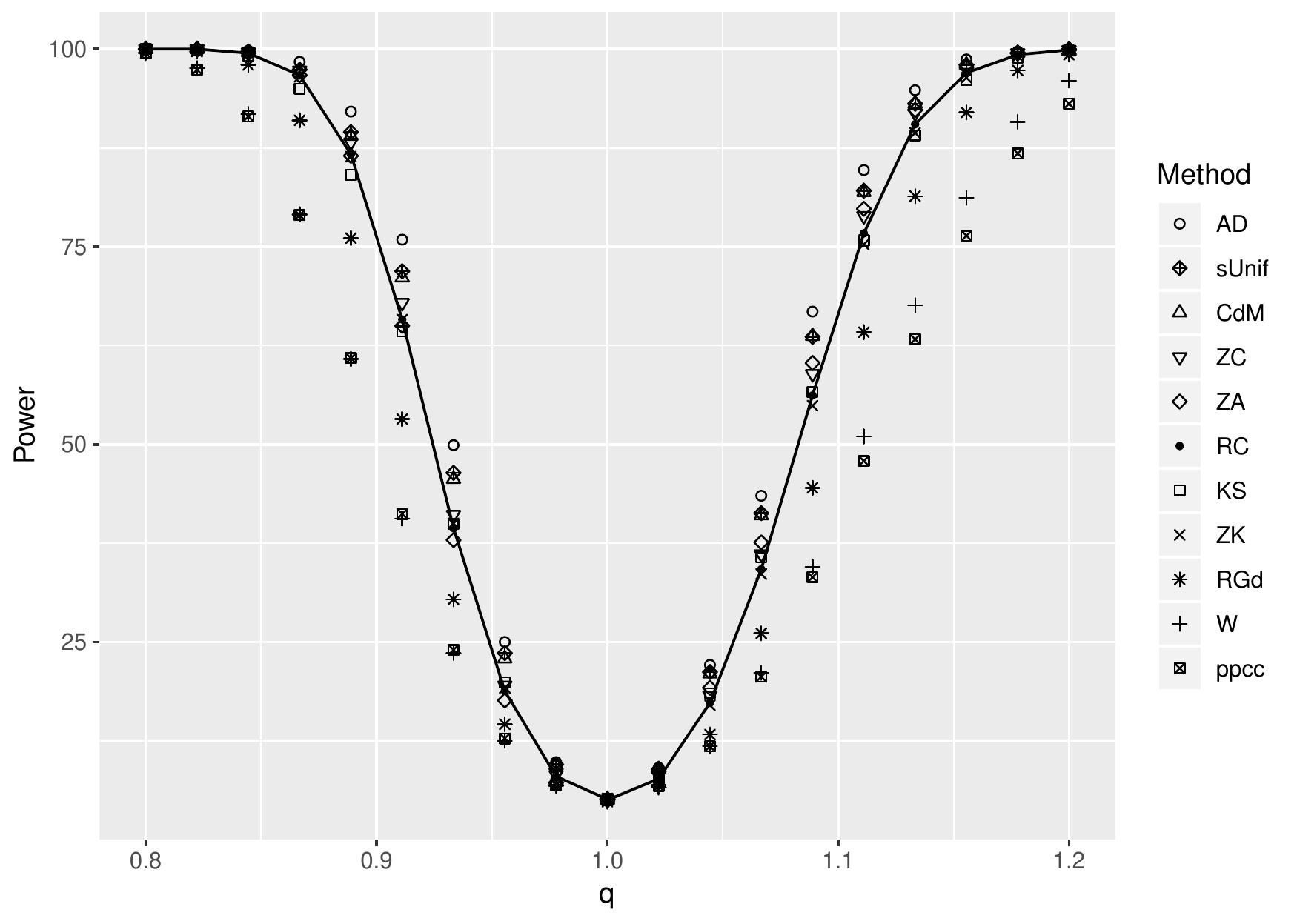}
\end{center}
\caption{Power of various tests if null hypothesis specifies a Uniform [0,1] and the true distribution is Beta(1,q).}
\label{fig:Case6}
\end{figure}

\subsubsection{U[0,1] vs Beta(1,q), Figure~\ref{fig:Case6}}

\clearpage
\newpage

\begin{figure}
\begin{center}
\includegraphics[width=4.5in]{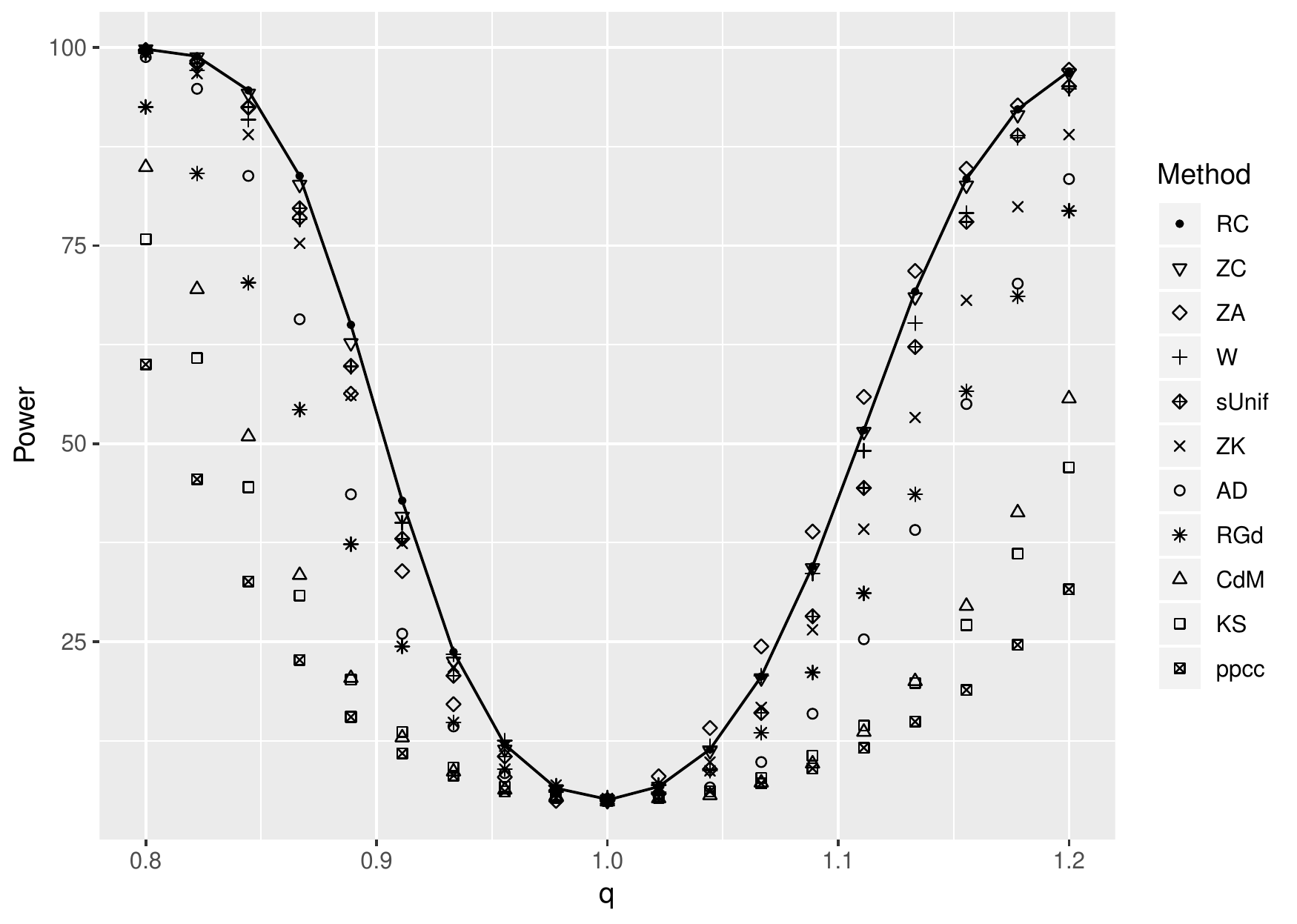}
\end{center}
\caption{Power of various tests if null hypothesis specifies a Uniform [0,1] and the true distribution is Beta(q,q).}
\label{fig:Case7}
\end{figure}

\subsubsection{U[0,1] vs Beta(q,q), Figure~\ref{fig:Case7}}

\clearpage
\newpage

\begin{figure}
\begin{center}
\includegraphics[width=4.5in]{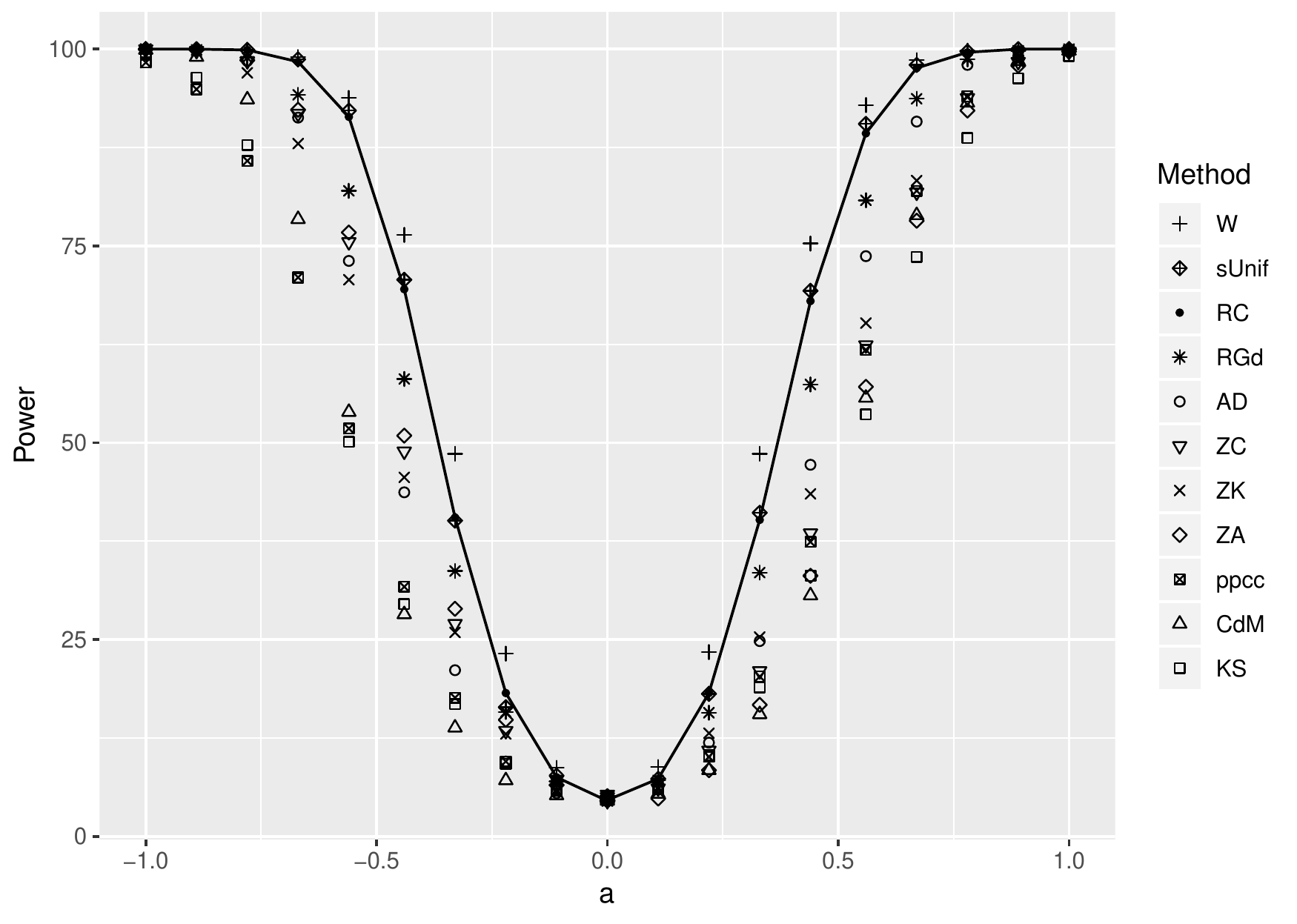}
\end{center}
\caption{Power of various tests if null hypothesis specifies a Uniform [0,1] and the true distribution is quadratic.}
\label{fig:Case8}
\end{figure}

\subsubsection{U[0,1] vs Quadratic, Figure~\ref{fig:Case8}}

The quadratic density is parametrized as $f(x;s)=3a(x-0.5)^2+1-a/4,0<x<1$, so the case $a=0$ corresponds to the U[0,1] distribution. 

\clearpage
\newpage

\begin{figure}
\begin{center}
\includegraphics[width=4.5in]{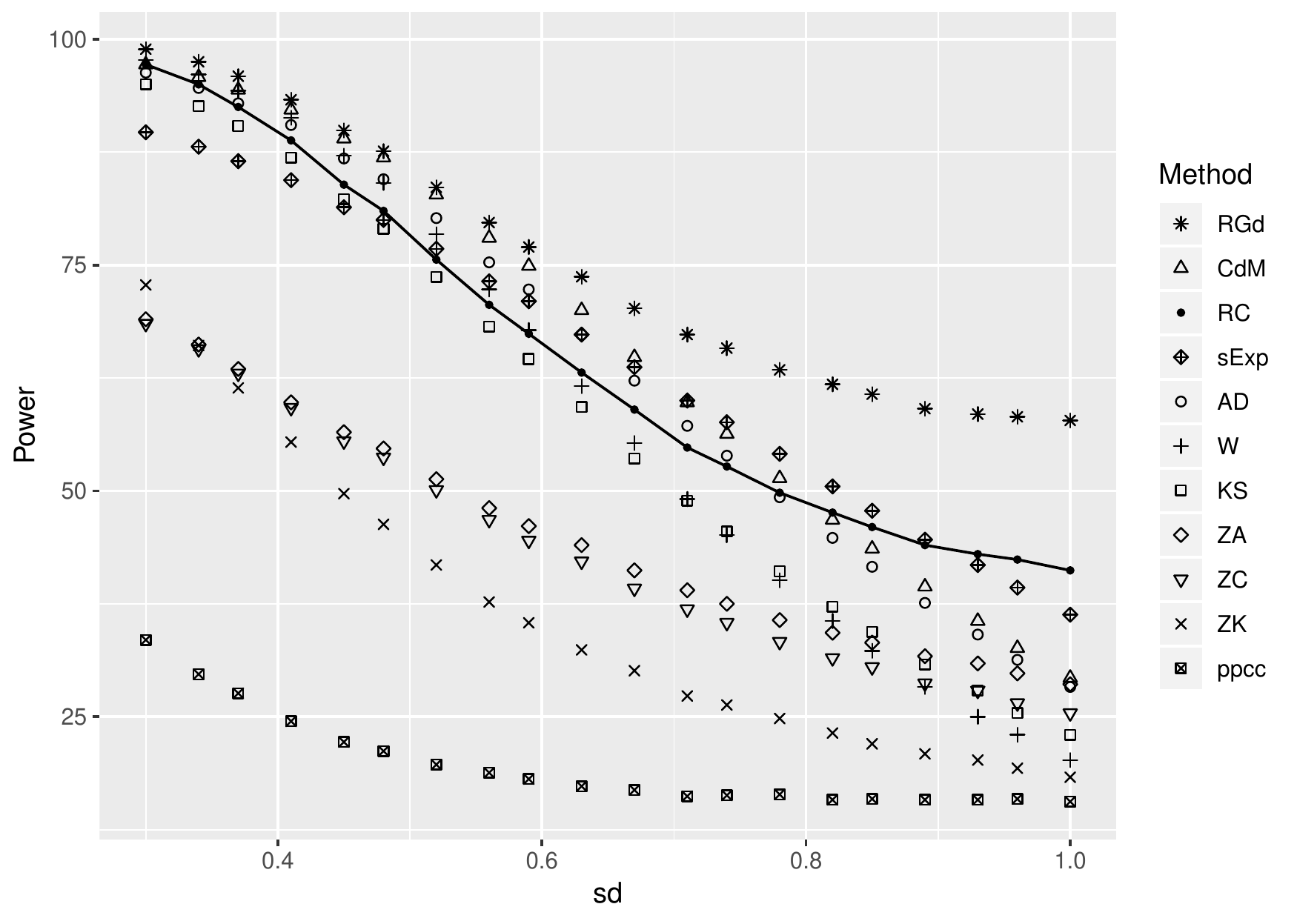}
\end{center}
\caption{Power of various tests if null hypothesis specifies an exponential distribution with estimated rate and the true distribution is exponential rate 1 with a normal bump mean 1.5 and varying standard deviaiton.}
\label{fig:Case9}
\end{figure}

The next four cases use the exponential distribution under the null hypothesis:

\subsubsection{Exponential vs Exponential(1)+Normal(1.5, $\sigma^2$), Figure~\ref{fig:Case9}}

Under the true distribution the density has a bump at 1.5. $\sigma=1:20:0.3$. The normal distribution is truncated to $x>0$. 

\clearpage
\newpage

\begin{figure}
\begin{center}
\includegraphics[width=4.5in]{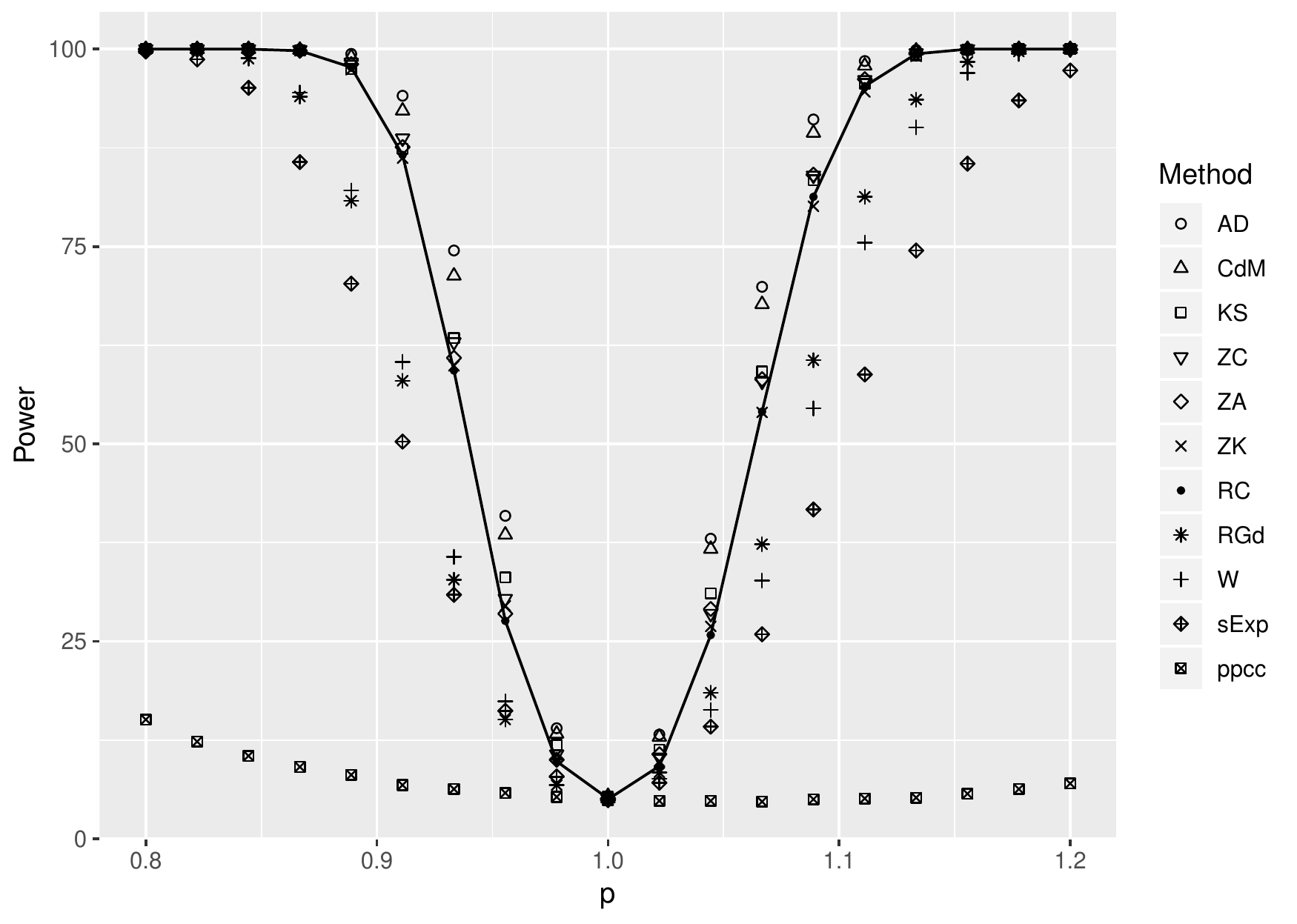}
\end{center}
\caption{Power of various tests if null hypothesis specifies an exponential distribution with rate 1 and the true distribution is Gamma(p, 1).} 
\label{fig:Case10}
\end{figure}

\subsubsection{Exponential(1) vs Gamma(p, 1), Figure~\ref{fig:Case10}}

\clearpage
\newpage

\begin{figure}
\begin{center}
\includegraphics[width=4.5in]{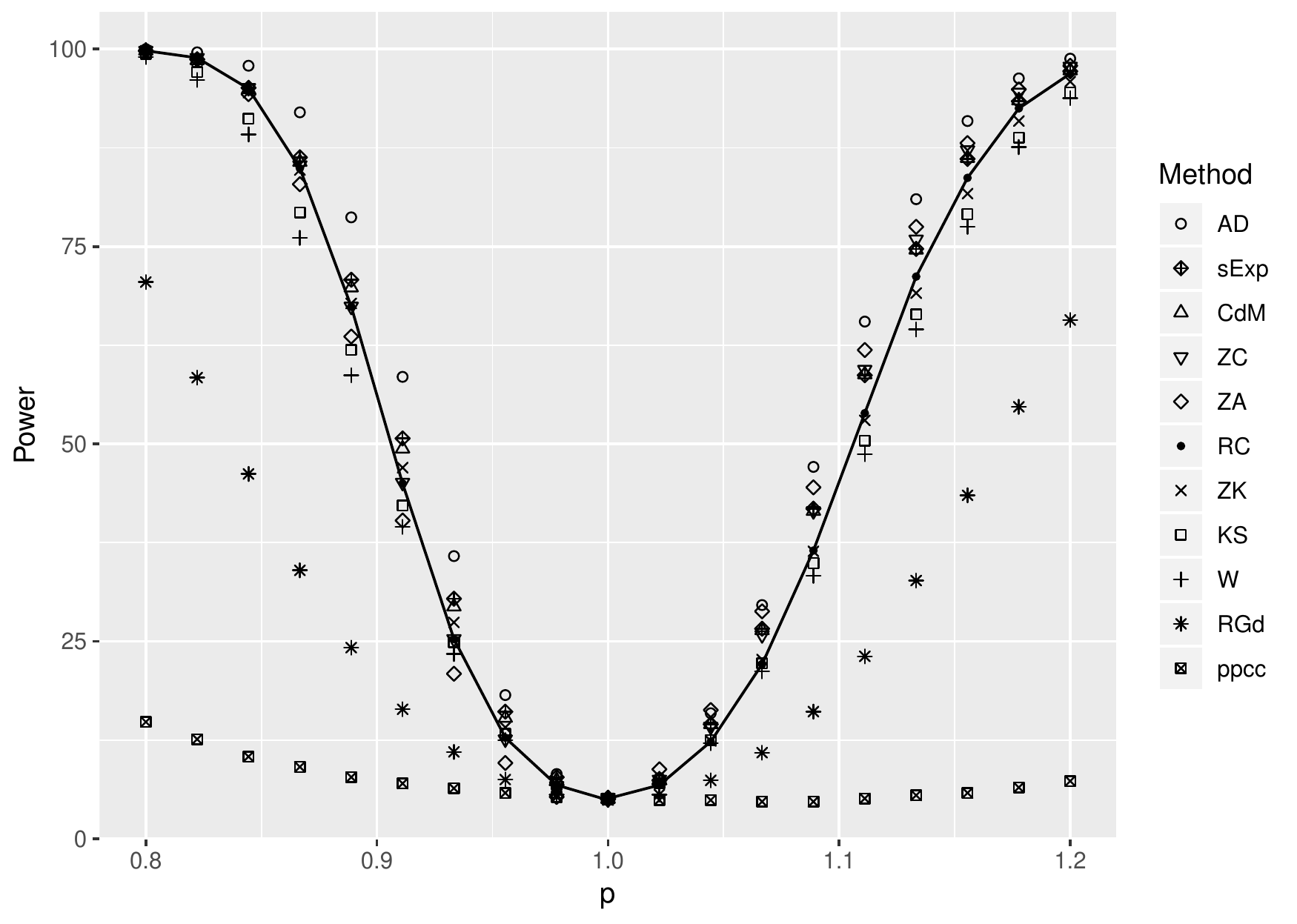}
\end{center}
\caption{Power of various tests if null hypothesis specifies an exponential distribution with rate estimated and the true distribution is Gamma(p, 1).}
\label{fig:Case11}
\end{figure}

\subsubsection{Exponential vs Gamma(p, 1), Figure~\ref{fig:Case11}}

The rate of the exponential is estimated via maximum likelihood.

\clearpage
\newpage

\begin{figure}
\begin{center}
\includegraphics[width=4.5in]{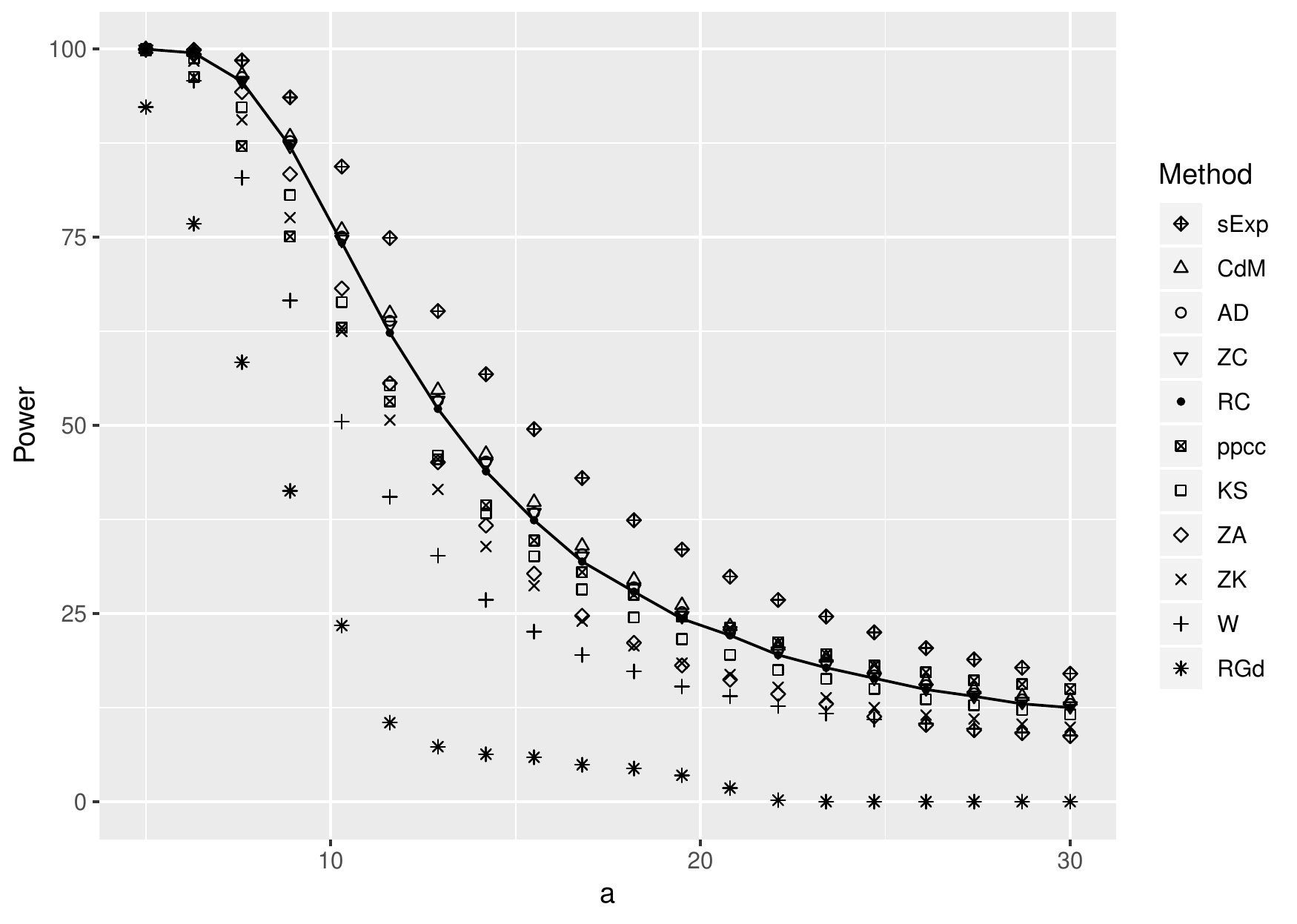}
\end{center}
\caption{Power of various tests if null hypothesis specifies an exponential distribution with rate estimated and the true distribution is Inverse.}
\label{fig:Case12}
\end{figure}

\subsubsection{Exponential vs Inverse Power, Figure~\ref{fig:Case12}}

The true density is parametrized as $f(x;a)=\frac{(a+1)}{(1+x)^a},x>0$. $a=5:20:30$.

\clearpage
\newpage

\begin{figure}
\begin{center}
\includegraphics[width=4.5in]{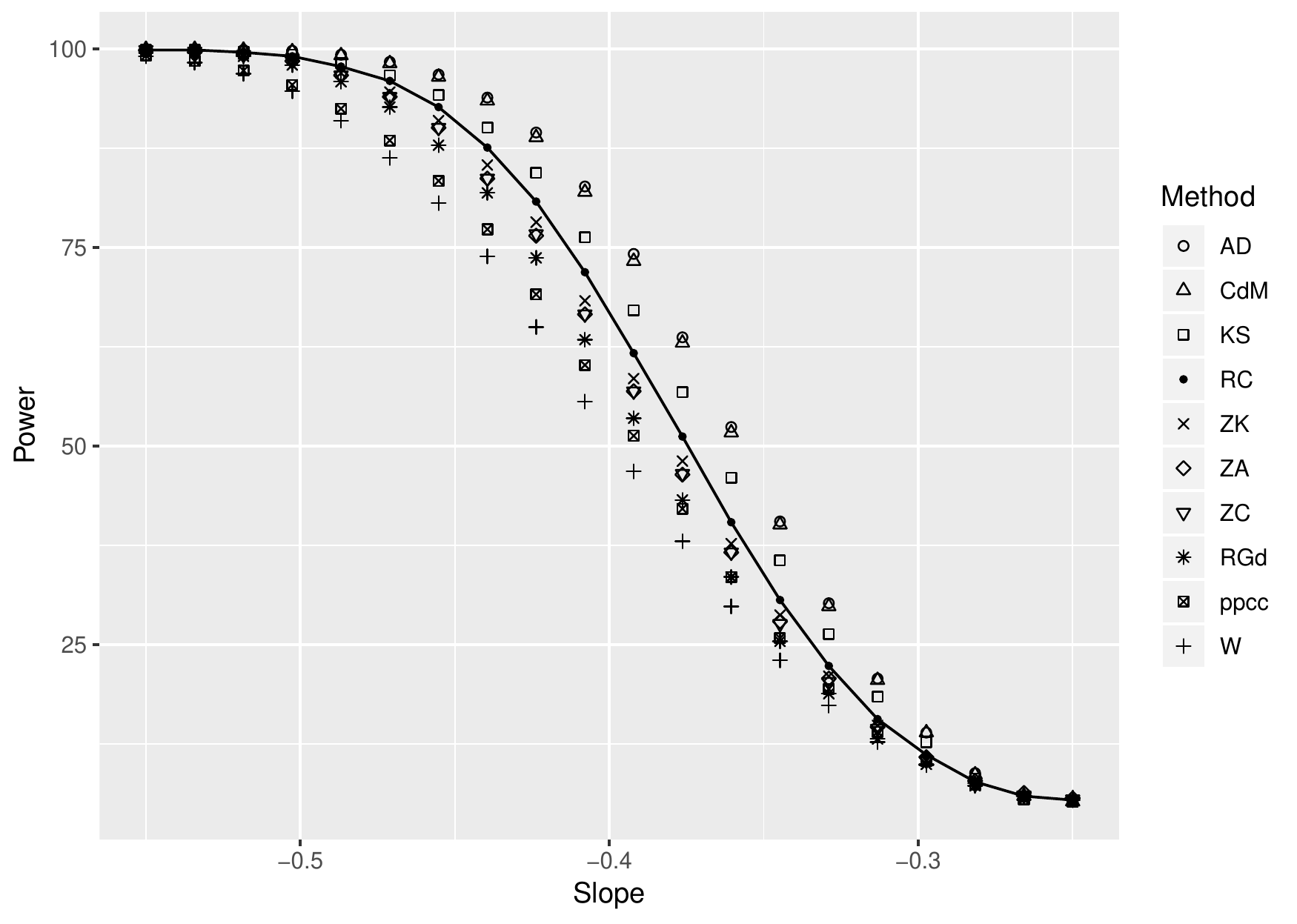}
\end{center}
\caption{Power of various tests if null hypothesis specifies an exponential distribution with rate 0.5 truncated to [0,1] and the true distribution is linear.}
\label{fig:Case13}
\end{figure}

\subsubsection{Truncated Exponential(0.5, 0, 1) vs Linear(p), Figure~\ref{fig:Case13}}

The null hypothesis specifies an exponential distribution rate 0.5, truncated to the interval $[0,1]$. The true distribution is a linear as above, with slope $s=-0.2:20:-0.5$. 

\clearpage
\newpage

\begin{figure}
\begin{center}
\includegraphics[width=4.5in]{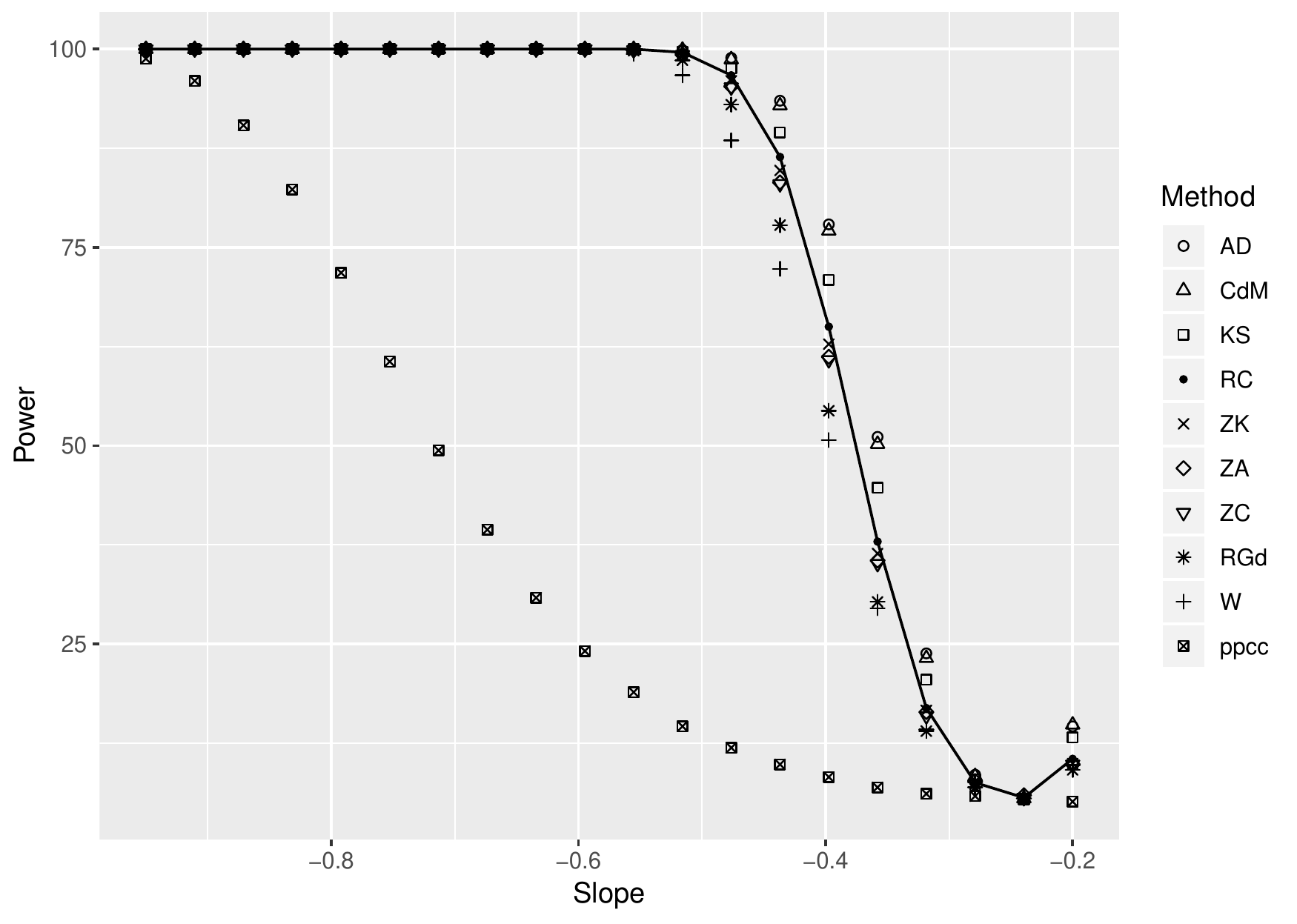}
\end{center}
\caption{Power of various tests if null hypothesis specifies an exponential distribution with estimated rate, truncated to [0,1], and the true distribution is linear.}
\label{fig:Case14}
\end{figure}

\subsubsection{Truncated Exponential(., 0, 1) vs Linear(p), Figure~\ref{fig:Case14}}

Same as last case, but now the rate is estimated via maximum likelihood.

\clearpage
\newpage

\begin{figure}
\begin{center}
\includegraphics[width=4.5in]{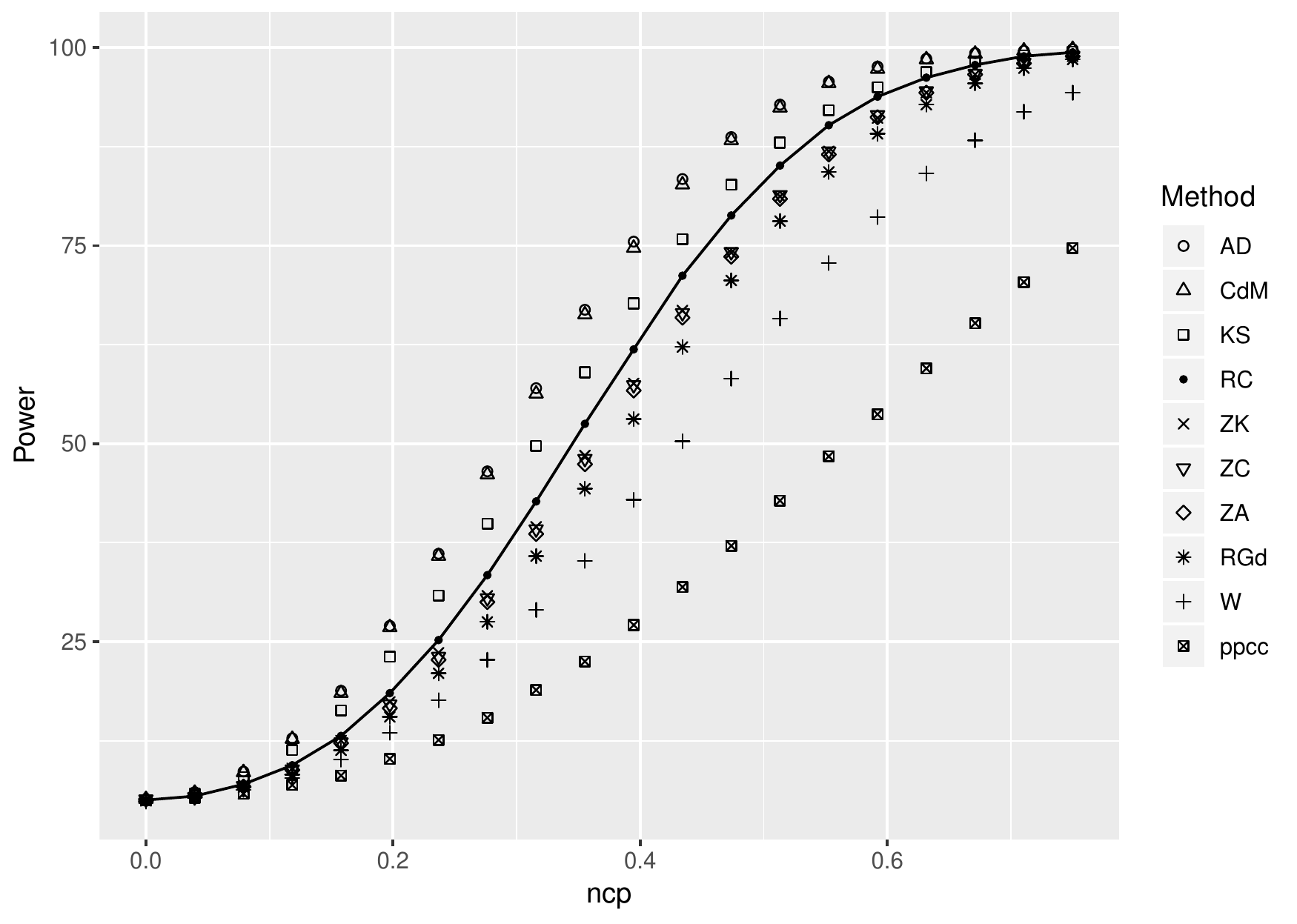}
\end{center}
\caption{Power of various tests if null hypothesis specifies a Beta( 2,2) and the true distribution is a non central Beta(2,2,p).}
\label{fig:Case15}
\end{figure}

\subsubsection{Beta(2,2) vs Beta(2,2,p), Figure~\ref{fig:Case15}}

The true distribution is a non-central Beta with non-centrality parameter $p=0:20:0.75$. 

\clearpage
\newpage

\begin{figure}
\begin{center}
\includegraphics[width=4.5in]{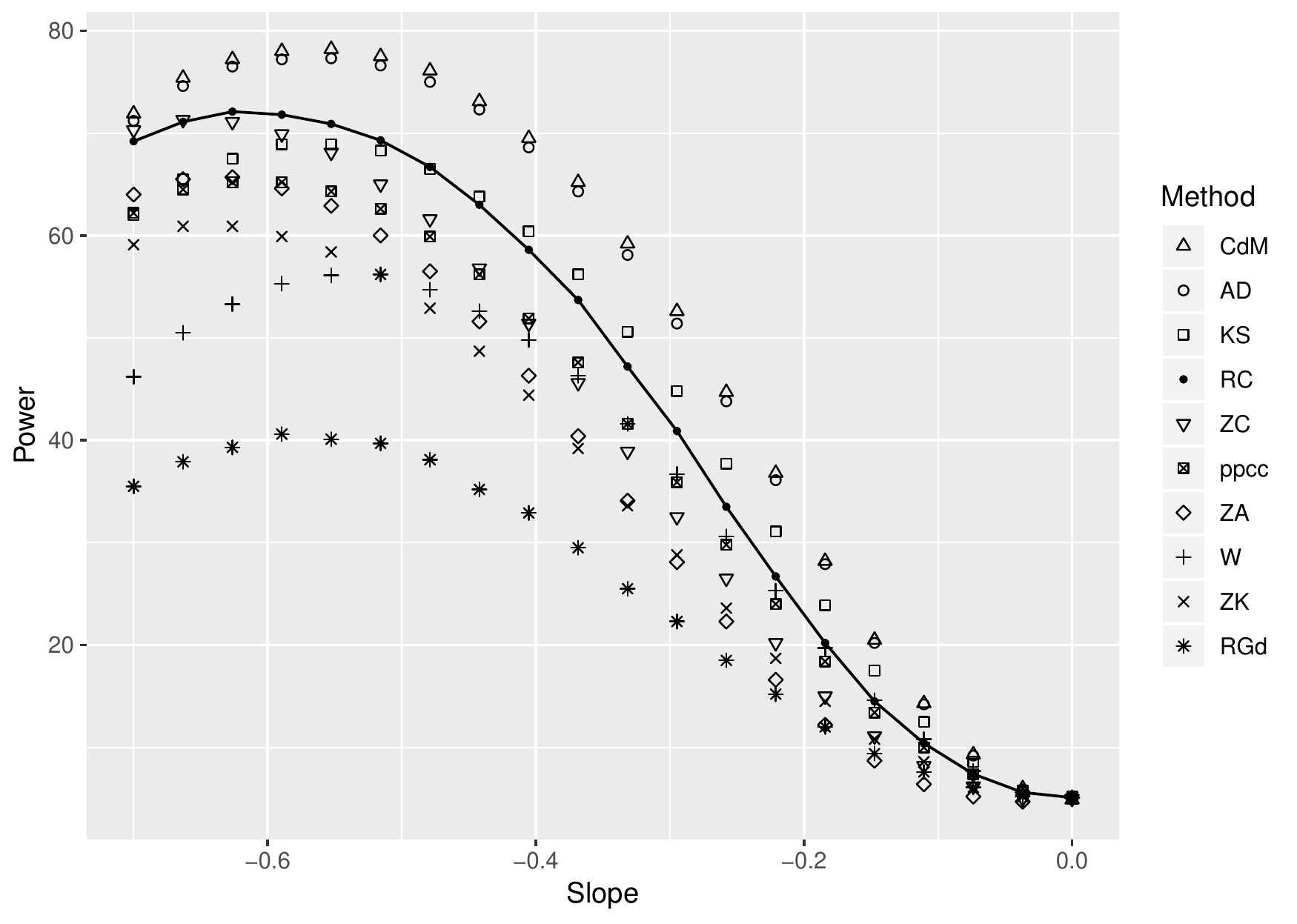}
\end{center}
\caption{Power of various tests if null hypothesis specifies a Beta (1, $\beta$) with $\beta$ estimated and the true distribution is linear.}
\label{fig:Case16}
\end{figure}

\subsubsection{Beta(1,.) vs Linear(s), Figure~\ref{fig:Case16}}

The null distribution is a Beta with $\alpha=1$ and $\beta$ estimated via maximum likelihood. 

\clearpage
\newpage

\begin{figure}
\begin{center}
\includegraphics[width=4.5in]{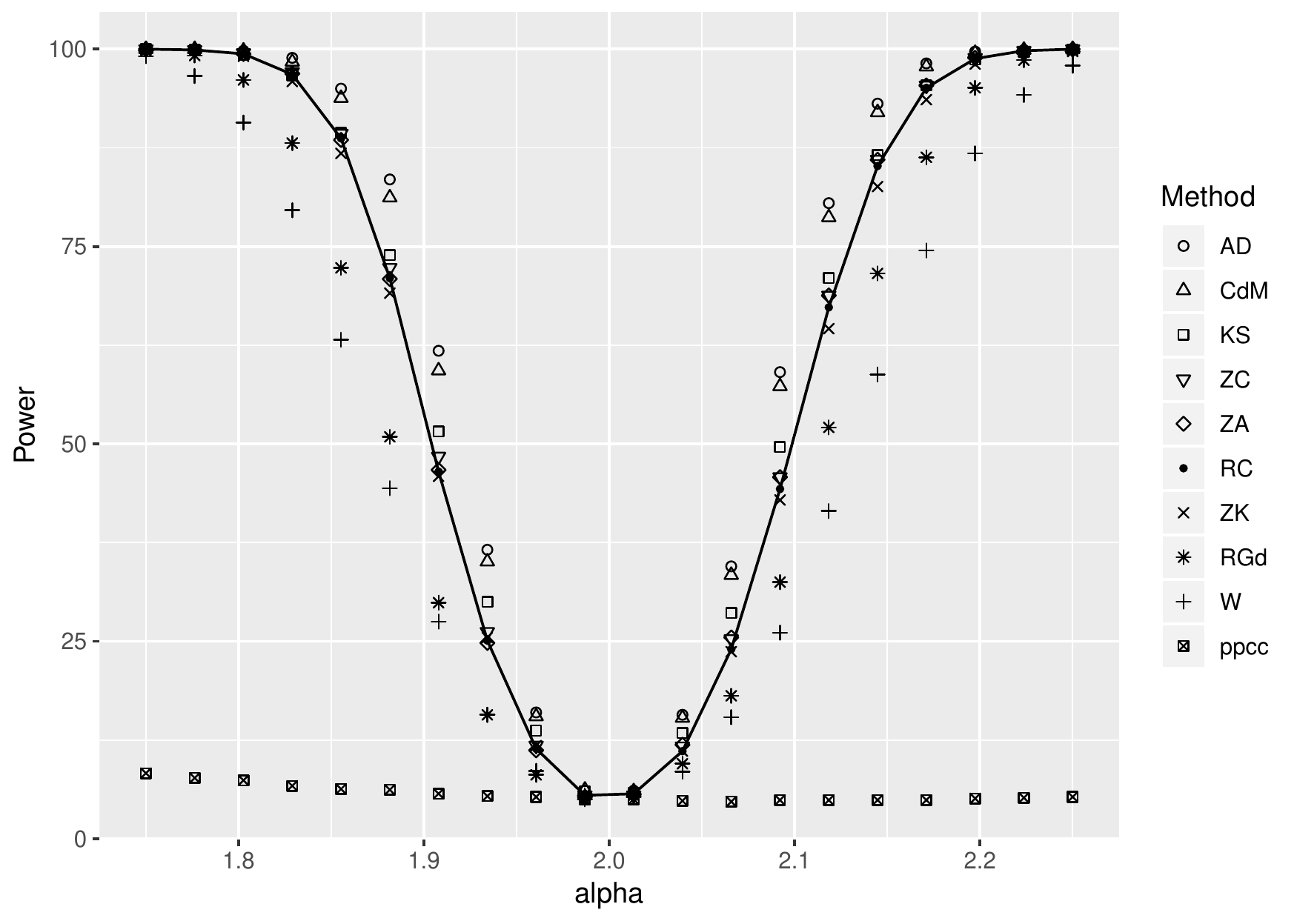}
\end{center}
\caption{Power of various tests if null hypothesis specifies an Erlang with parameters estimated and the true distribution is Gamma($\alpha$, 5).}
\label{fig:Case17}
\end{figure}

\subsubsection{Erlang(., .) vs Gamma($\alpha$,5), Figure~\ref{fig:Case17}}

The null distribution is Erlang with the parameters estimated via method of moments. The true distribution is Gamma($\alpha$,5), where $\alpha=1.75:20:2.25$. 

\clearpage
\newpage

\begin{figure}
\begin{center}
\includegraphics[width=4.5in]{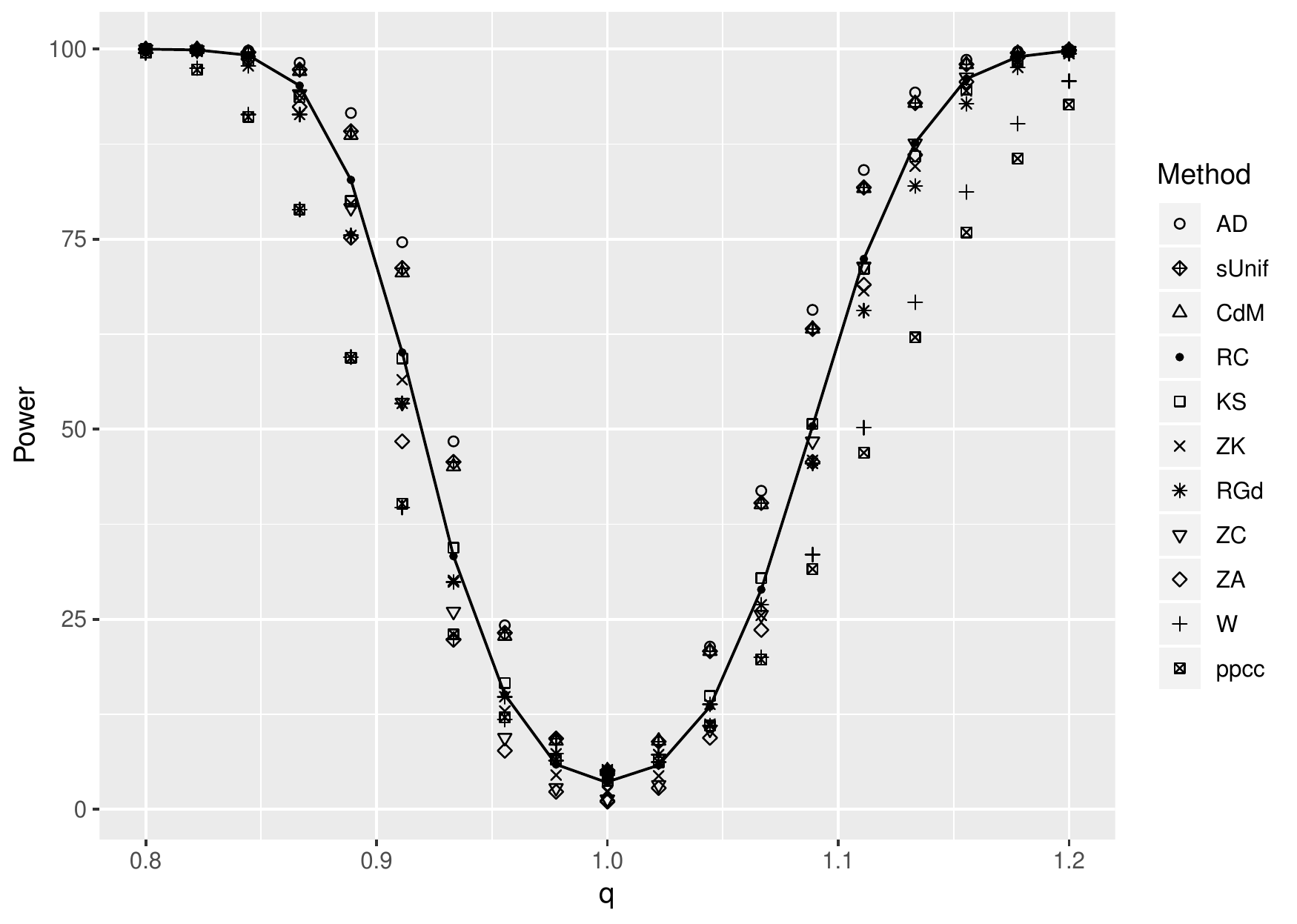}
\end{center}
\caption{Power of various tests if null hypothesis specifies a Uniform[0,1], the true distribution is Beta(1,q) and the data is binned.}
\label{fig:Case18}
\end{figure}

\subsubsection{Uniform[0,1] vs Beta(1, q), binned data, Figure~\ref{fig:Case18}}

Here the data is in the form of a histogram with 50 equal-sized bins. $q=0.8:20:1.2$.

\clearpage
\newpage

\begin{figure}
\begin{center}
\includegraphics[width=4.5in]{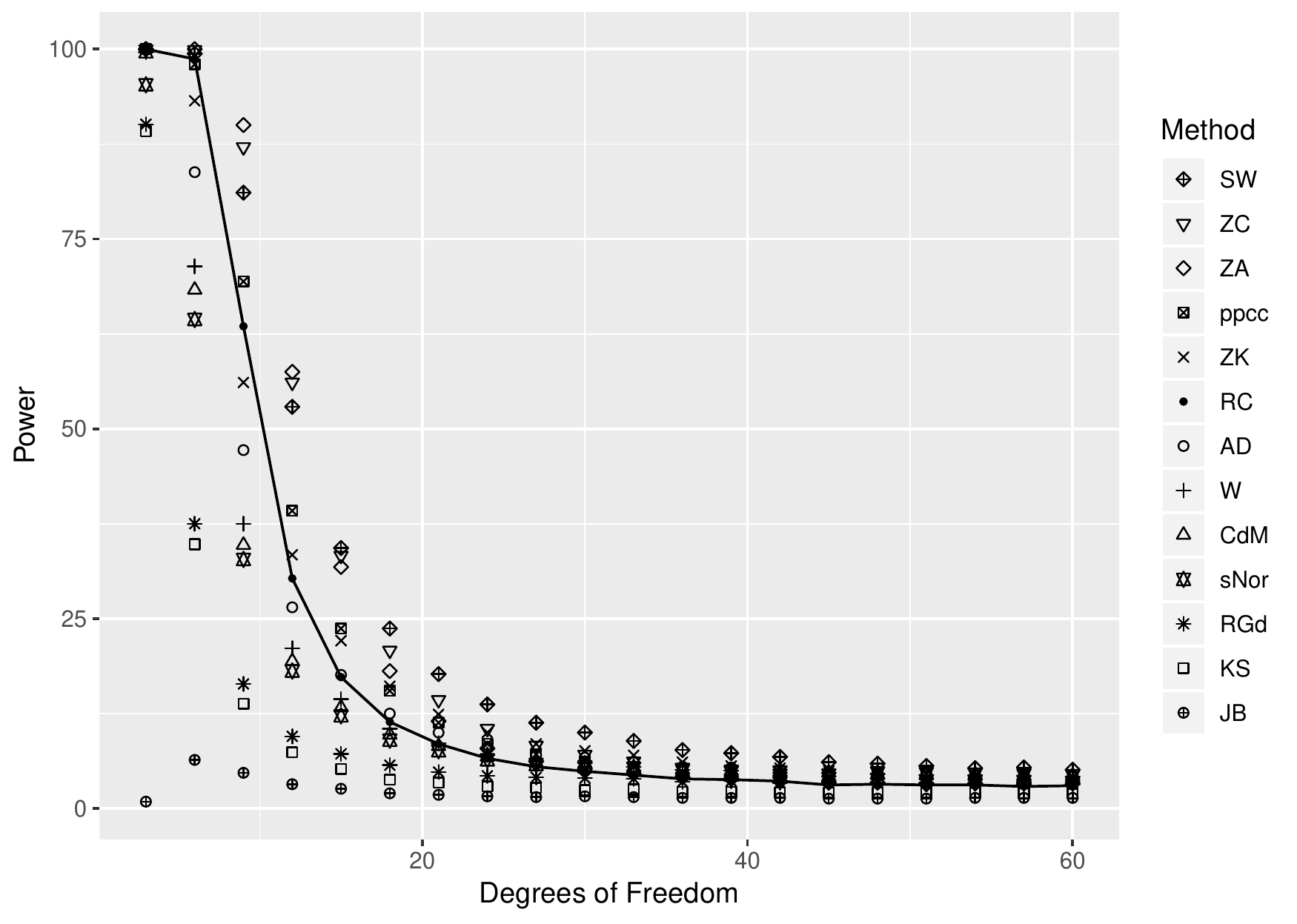}
\end{center}
\caption{Power of various tests if null hypothesis specifies a normal distribution with ean and standard deviation estimated, the true distribution is t and the data is binned.}
\label{fig:Case19}
\end{figure}

\subsubsection{Normal vs t(n), binned data, Figure~\ref{fig:Case19}}

Again the data is in the form of a histogram with 50 bins. The mean and standard deviation are estimated via maximum likelihood.

\clearpage
\newpage

\begin{figure}
\begin{center}
\includegraphics[width=4.5in]{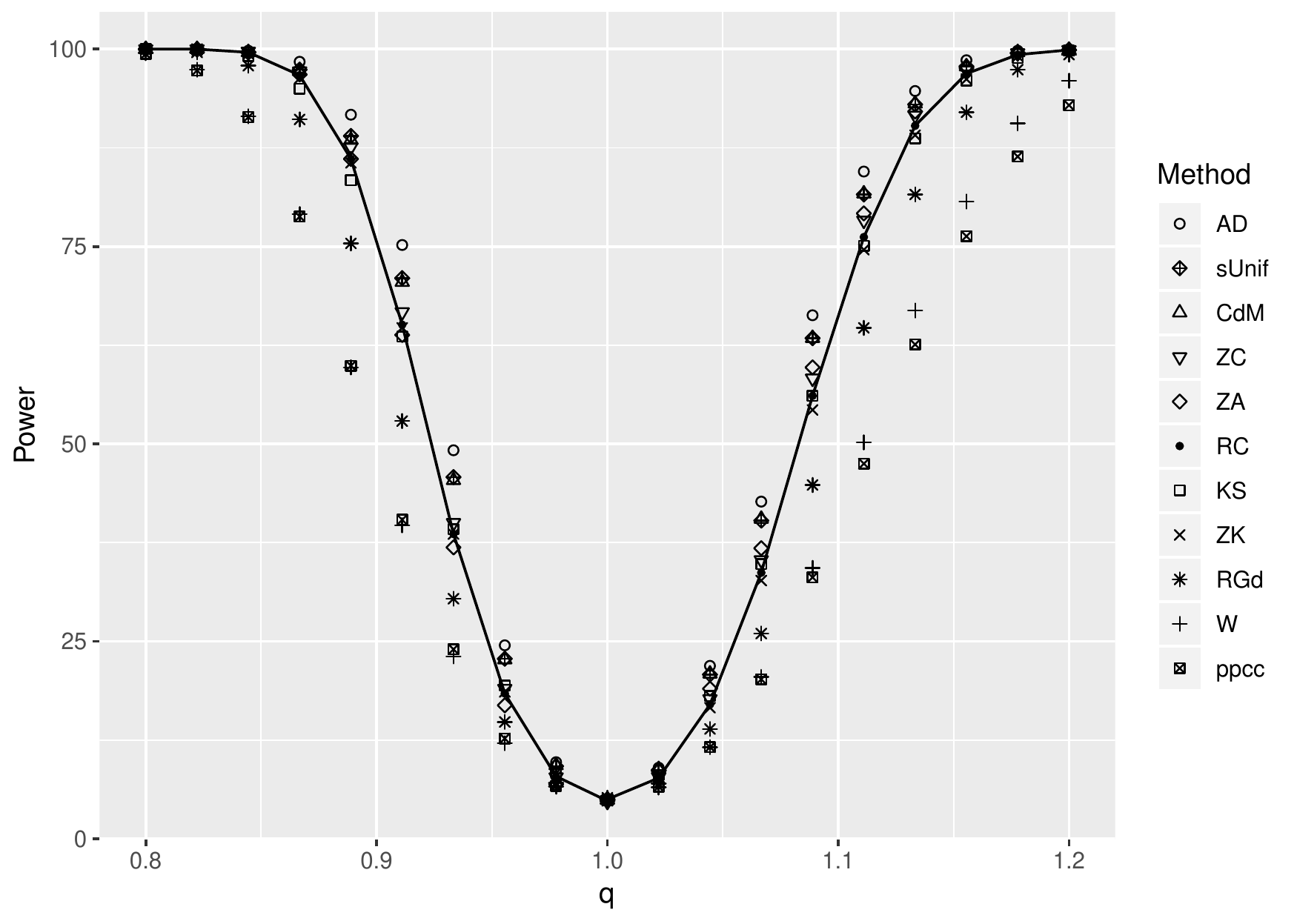}
\end{center}
\caption{Power of various tests if null hypothesis specifies a Uniform [0,1], the true distribution is Beta(1,q) an the sample size is drawn from a Poisson(1000).}
\label{fig:Case20}
\end{figure}

\subsubsection{Uniform[0,1] vs Beta(1, q), Poisson sample size, Figure~\ref{fig:Case20}}

In this case the sample size varies according to a Poisson random variable with rate $\lambda=1000$..

\clearpage
\newpage

\subsection{Overall Performance}

In this section we compare the performance of the various methods. Note that in almost all cases if method A had higher power than method B for one value of the parameter, it did so for all of them. It is therefore reasonable to compare their mean powers. 

\subsubsection{Mean Power}

Our case studies include 21 different null hypotheses and true distributions, each with 20 different parameter values for a total of 420 cases. If we simply find the mean power of the methods used in all simulations we find that RC has the highest mean power at 49.18\%, followed by ZC (48.91\%), AD (48.77\%), ZA (48.20\%), CdM (46.08\%), ZK (45.97\%), KS (42.66\%), W (42.25\%), RGd (41.67\%) and finally ppcc (35.27\%). So the method proposed in this paper achieves the highest average power over the 21 case studies. While any study of this kind is necessarily limited, this does suggest that the RC method performs quite well.

\subsubsection{Rankings of Methods by Case}

\begin{table}
\centering
\caption{Ranking of methods for each case study}
\begin{tabular}{l|r|r|r|r|r|r|r|r|r|r}
\hline
  & RC & ZC & AD & ZA & CdM & ZK & KS & W & RGd & ppcc\\
\hline\hline
Case 1 & 9 & 8 & 5 & 6 & 3 & 7 & 2 & 4 & 1 & 10\\
\hline
Case 2 & 9 & 10 & 4 & 8 & 2 & 7 & 1 & 3 & 5 & 6\\
\hline
Case 3 & 8 & 9 & 6 & 10 & 3 & 5 & 2 & 4 & 1 & 7\\
\hline
Case 4 & 9 & 7 & 3 & 8 & 1 & 6 & 2 & 4 & 5 & 10\\
\hline
Case 5 & 9 & 7 & 5 & 8 & 4 & 6 & 3 & 1 & 2 & 10\\
\hline
Case 6 & 7 & 5 & 9 & 3 & 10 & 6 & 8 & 1 & 2 & 4\\
\hline
Case 7 & 6 & 8 & 10 & 7 & 9 & 4 & 5 & 2 & 3 & 1\\
\hline
Case 8 & 10 & 9 & 5 & 8 & 3 & 6 & 2 & 7 & 4 & 1\\
\hline
Case 9 & 9 & 6 & 7 & 4 & 2 & 5 & 1 & 10 & 8 & 3\\
\hline
Case 10 & 8 & 3 & 7 & 4 & 9 & 2 & 5 & 6 & 10 & 1\\
\hline
Case 11 & 4 & 7 & 10 & 6 & 9 & 5 & 8 & 2 & 3 & 1\\
\hline
Case 12 & 6 & 8 & 10 & 7 & 9 & 5 & 4 & 3 & 2 & 1\\
\hline
Case 13 & 7 & 8 & 9 & 4 & 10 & 3 & 5 & 2 & 1 & 6\\
\hline
Case 14 & 7 & 4 & 10 & 5 & 9 & 6 & 8 & 1 & 3 & 2\\
\hline
Case 15 & 7 & 4 & 10 & 5 & 9 & 6 & 8 & 2 & 3 & 1\\
\hline
Case 16 & 7 & 5 & 10 & 4 & 9 & 6 & 8 & 2 & 3 & 1\\
\hline
Case 17 & 7 & 6 & 9 & 4 & 10 & 2 & 8 & 3 & 1 & 5\\
\hline
Case 18 & 5 & 7 & 10 & 6 & 9 & 4 & 8 & 2 & 3 & 1\\
\hline\hline
Mean Rank & 7.4 & 6.7 & 7.7 & 5.9 & 6.7 & 5.1 & 4.9 & 3.3 & 3.3 & 3.9\\
\end{tabular}
\end{table}

Another way to study the performance of the methods is as follows: for each of the 18 null hypotheses (excluding the special cases such as binned data) we rank the methods, with a rank of 1 for the method with the highest mean power. Next we find the number of times a method had rank 1, rank 2 and so on.

The results are shown in table 2, together with the mean rankings over all cases. The RC method has the second highest mean ranking after Anderson-Darling. While Anderson-Darling is certainly a very good method, and would be our choice if only a single method can be used, it also is likely to perform badly in some cases. While we did not find such a case it almost certainly exists. On the other hand by its design RC should never perform especially badly. 

\begin{figure}
\begin{center}
\includegraphics[width=4in]{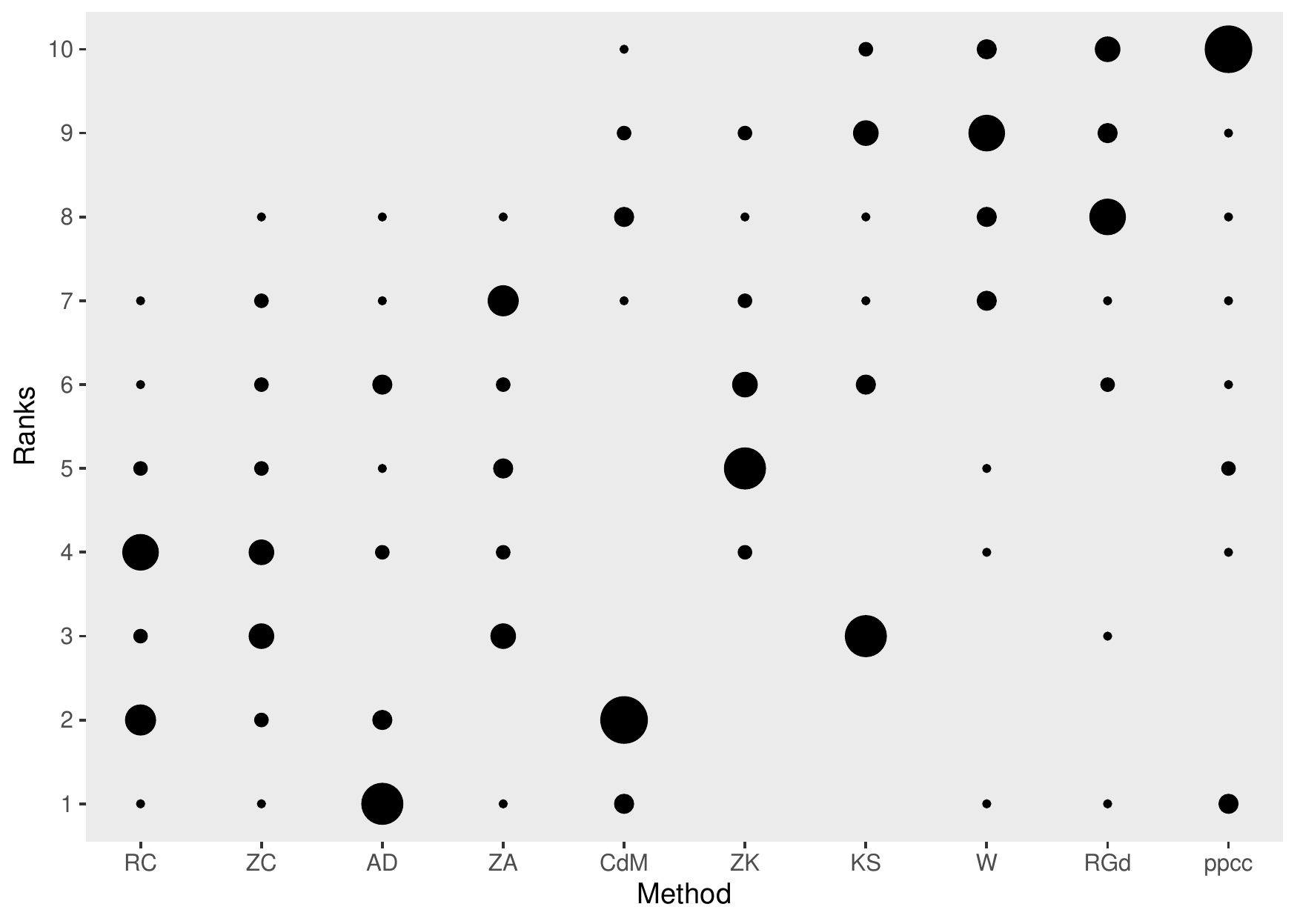}
\end{center}
\caption{Number of times each method was best, had rank 2 and so on. The size of the plotting symbol indicates the frequency of each rank, 1 being best.} 
\label{fig:Result1}
\end{figure}

Figure~\ref{fig:Result1} also illustrates the rankings, with the methods sorted by their overall mean power. The frequency a rank was attained is indicated by the size of the plotting symbol. RC was ranked best once, second five times, third twice etc. On the other hand RC was never worse than seventh. Four methods (CdM, W, RGd and ppcc) were best in some cases and worst in others, indicating the difficulty to choose a single method.

\subsubsection{Difference in Power to Best Method}

\begin{figure}
\begin{center}
\includegraphics[width=4in]{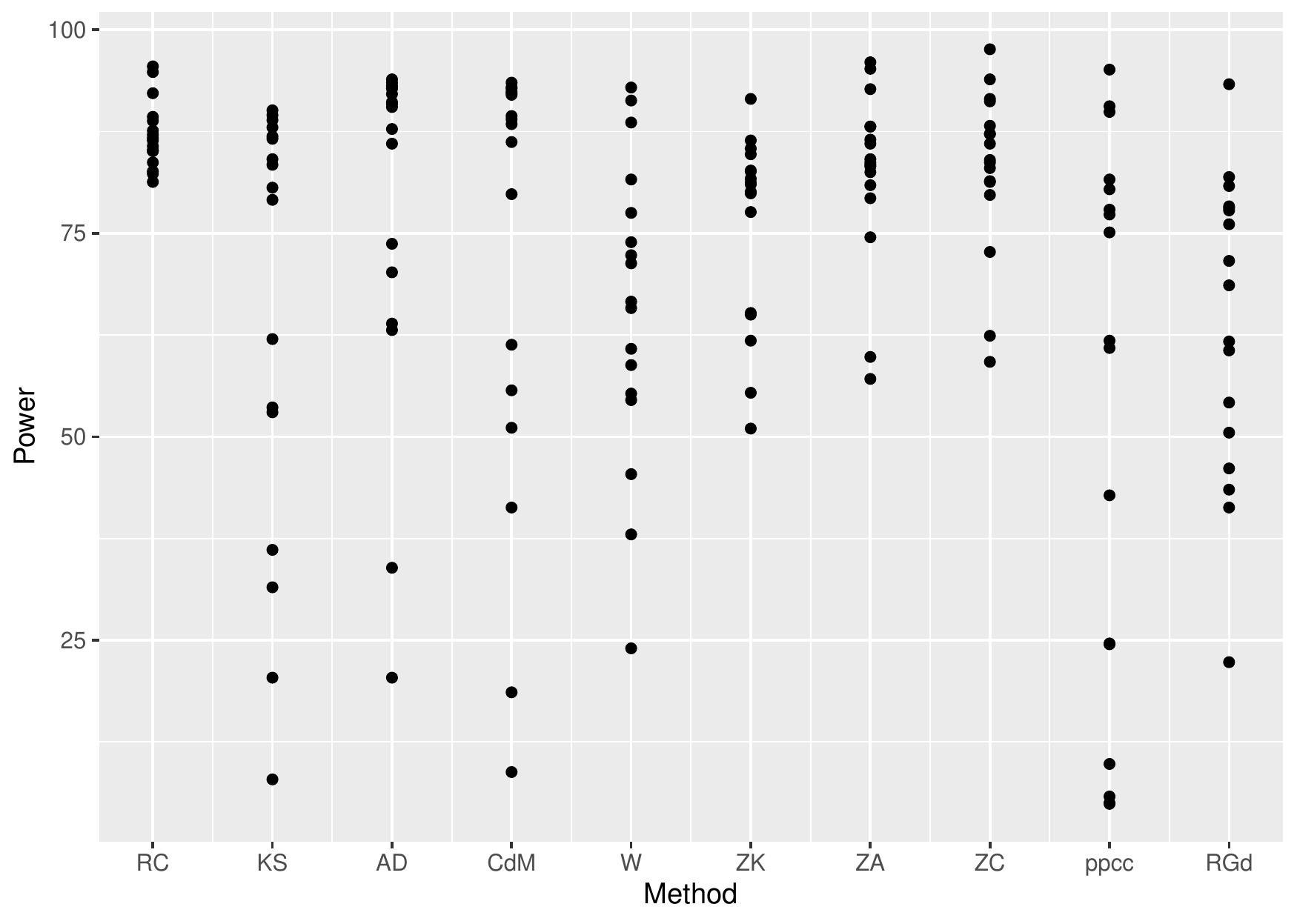}
\end{center}
\caption{Power of each method when best one has a power just above 90\%.}
\label{fig:Result2}
\end{figure}

Finally we consider for each case how much lower the power of each method is when compared to 
the best. To do so we find for each case the value of the parameter where at least one method 
has a power just over 90\%. For this value of the parameter all powers are recorded. They are 
shown in Figure~\ref{fig:Result2}. The power of RC is never less than 80\%, or about 10\% 
below the best method whereas the individual methods sometimes perform much worse. Even Anderson-Darling had a power below $50\%$ twice. So using RC 
guards against ever having exceptionally low power.

\section{Computational Issues}

An R library as well as the source code with all necessary routines to carry out this test is available from https://github.com/WolfgangRolke/simgof. 
Alternatively it can be run as an R shiny app at https://drrolke.shinyapps.io/sgoftest. The app allows the user to upload the data set and 
 define the necessary routines (to calculate the distribution function, generate new data from the distribution for the Monte Carlo, optionally calculate the quantile function and do parameter estimation) either using R or C++ code. The code to run the app locally is also available at https://github.com/WolfgangRolke/simgof and detailed explanations on how to run the app can be found at  http://academic.uprm.edu/wrolke/simgof.explained.pdf. 

\section{Conclusion}  
 
We presented a method that combines several standard goodness-of-fit tests for continuous 
distributions. The test rejects the null hypothesis at the $\alpha$ level if any of the 
individual tests does. Simulation is used to adjust the p value so it has a uniform [0,1] 
distribution. Extensive simulation studies show that this method does indeed achieve 
the desired nominal type I error probability and  when averaged over the 21 cases included 
in this study has overall power better than any of the individual tests. While some methods such as
Anderson-Darling and Zhang ZC generally perform well they also can have quite low power whereas RC should 
never be much worse than the best method, simply because that best method is part of RC.

\bibliographystyle{Chicago}

\bibliography{simultaneous-gof-test}

\section{Appendix - Source Code}

In the following we have the source code for the routines used.

\section{TS}

This routine calculates the test statistics for all the tests included.

\begin{lstlisting}

TS <- function(x, case) {
  if(is.list(x))
    x <- simgof::spreadout(x, case)
  # data is binned, unbin it
  n <- length(x)
  x <- sort(x)
  param <- case$est.mle(x)
  y <- case$pnull(x, param)
  m <- 1:n-0.5
  out <- rep(0, length(case$methods))
  names(out) <- case$methods
  tmp <- c(  max(c(y-0:(n-1)/n, 1:n/n-y)),
            -n-mean((2*1:n-1)*(log(y)+log(1-y[n:1]))),
            1/(12*n)+sum( ((2*(1:n)-1)/2/n- y)^2 ),
            1/(12*n)+sum( ((2*(1:n)-1)/2/n- y)^2 )-n*(mean(y)-0.5)^2,
            max(m*log(m/n/y)+(n-m)*log((n-m)/n/(1-y))),
            (-1)*sum(log(y)/(n-m)+log((1-y))/m),
            sum(log( (1/y-1)/((n-0.5)/(1:n-0.75)-1)  )^2))
  names(tmp) <- c("KS", "AD", "CdM", "W", "ZK", "ZA", "ZC")
  for(m in case$methods) {
    if(m %in% c("KS", "AD", "CdM", "W", "ZK", "ZA", "ZC"))
      out[m] <- tmp[m]
  }
  if("SW" %in% case$methods)
    out["SW"] <- 1-shapiro.test(x)$statistic
  if("ppcc" %in% case$methods)
    out["ppcc"] <- 1-cor(x, case$qnull(ppoints(case$n), param))
  if("JB" %in% case$methods) {
    mu <- mean(x)
    S <- mean((x-mu)^3)/(mean((x-mu)^2))^(3/2)
    K <- mean((x-mu)^4)/(mean((x-mu)^2))^2
    out["JB"] <- n/6*(S^2+(K-3)^2/4)
  }
  if("RGd" %in% case$methods) {
    out["RGd"] <- chisquare.test(x, case, "RGd")
  }
  if("Equal Size" %in% case$methods) {
    out["Equal Size"] <- chisquare.test(x, case, "Equal Size")
  }
  if("Equal Prob" %in% case$methods) {
    out["Equal Prob"] <- chisquare.test(x, case, "Equal Prob")
  }
  if("sNor" %in% case$methods) {
    out["sNor"] <- ddst::ddst.norm.test(x, compute.p = FALSE)$statistic
  }
  if("sUnif" %in% case$methods) {
    out["sUnif"] <- ddst::ddst.uniform.test(x, compute.p = FALSE)$statistic
  }
  if("sExp" %in% case$methods) {
    out["sExp"] <- ddst::ddst.exp.test(x, compute.p = FALSE)$statistic
  }
  out
}
\end{lstlisting}

\section{simgof.test}

This routine runs the test

\begin{lstlisting}

simgof.test <- function(x, pnull, rnull, qnull=function(x) NULL, 
    do.estimation=TRUE, estimate = function(x) NULL,                     
    include.methods = c(rep(TRUE, 7), rep(FALSE, 9)),    
    B=1000, lambda) {
  methods <- c("KS", "AD", "CdM",  "W", "ZA", "ZK",  "ZC",
           "RGd", "Equal Size", "Equal Prob", 
           "ppcc", "JB", "SW", "sNor", "sUnif", "sExp")
# step 1: do some setup work
    param <- NULL
    if(do.estimation) param <- estimate(x)
    case <- list(B=B, 
               param = param,
               methods = methods[include.methods],
               n = ifelse(is.list(x), sum(x$counts), length(x)),
               pnull = ifelse(do.estimation, pnull, function(x, param=1) pnull(x)), 
               rnull = ifelse(do.estimation, rnull, function(n, param=1) rnull(n)),
               qnull = ifelse(do.estimation, qnull, function(x, param=1) qnull(x)),
               est.mle = estimate,
               dta = x
               )
# step 2: find null distributions of each test
  znull <- matrix(0, B, length(case$methods))
  colnames(znull) <- case$methods
  for(i in 1:B) {
    case$n <- ifelse(missing(lambda), case$n, rpois(1, case$lambda))
    znull[i, ] <- simgof::TS(case$rnull(case$n, case$param), case)
  }
# step 3: find p values for each test, find their minimum
  tmp <- rep(0, length(case$methods))
  names(tmp) <- case$methods
  pval <- rep(0, case$B)
  for(i in 1:case$B) {
    xsim <- znull[sample(1:B, 1), ]
    for(k in case$methods) 
      tmp[k] <- sum(xsim[k]<znull[, k])/case$B      
    pval[i] <- min(tmp)
  }
# step 4: find cdf of p values
  x <- seq(0, 1, length=250)
  y <- 0*x
  for(i in 1:250) y[i] <- sum(pval<=x[i])/length(pval)
  xy <- cbind(x, y)
  adjust <- function(xy, a) {
    approx(x=xy[, 1], y=xy[, 2], xout=a, rule=2)$y
  }
# step 5: run test on data  
  TS.data <- simgof::TS(case$dta, case)
  pvals <- rep(0, length(case$methods))
  names(pvals) <- case$methods
  for(k in case$methods)
    pvals[k] <- sum(TS.data[k]<znull[, k])/case$B
  pvals <- c(adjust(xy, min(pvals)), pvals)
  names(pvals)[1] <- "RC"
  round(pvals, 4)
}

\end{lstlisting}

\section{chisquare.test}

This routine runs the chisquare test, if desired

\begin{lstlisting}

chisquare.test <- function (x, case, which="RGd") {
  bin.fun <- function (case, k, kappa) {
    n <- case$n
    L <- min(case$dta)
    R <- max(case$dta)
    bins0 <- c(L, case$qnull((1:(k - 1))/k), R)
    if (k == 2) 
      bins1 <- c(L, case$qnull(0.5), R)
    else {
      if (is.finite(L) & is.finite(R)) 
        bins1 <- seq(L, R, length = k + 1)
      if (is.finite(L) & !is.finite(R)) {
        R <- case$qnull(1 - 5/n)
        bins1 <- c(seq(L, R, length = k), Inf)
      }
      if (!is.finite(L) & is.finite(R)) {
        L <- case$qnull(5/n)
        bins1 <- c(-Inf, seq(L, R, length = k))
      }
      if (!is.finite(L) & !is.finite(R)) {
        L <- case$qnull(5/n)
        R <- case$qnull(1 - 5/n)
        bins1 <- c(-Inf, seq(L, R, length = k - 1), Inf)
      }
    }
    bins <- (1 - kappa) * bins0 + kappa * bins1
    if (is.nan(bins[1])) 
      bins[1] <- (-Inf)
    if (is.nan(bins[k + 1])) 
      bins[k + 1] <- Inf
    bins <- bin.adjust(case, bins)
    bins
  }
  bin.adjust <- function (case, bins) {
    p <- case$param
    E <- case$n * diff(case$pnull(bins, p))
    if (all(E > 5)) return(bins)
    nbins <- length(E)
    repeat {
      k <- which.min(E)
      if (k == 1) {
        bins <- bins[-2]
        E[1] <- E[1] + E[2]
        E <- E[-2]
      }
      if (k == nbins) {
        bins <- bins[-nbins]
        E[nbins - 1] <- E[nbins] + E[nbins - 1]
        E <- E[-nbins]
      }
      if (k > 1 & k < nbins) {
        if (E[k - 1] < E[k + 1]) {
          bins <- bins[-k]
          E[k] <- E[k] + E[k - 1]
          E <- E[-k]
        }
        else {
          bins <- bins[-(k + 1)]
          E[k] <- E[k] + E[k + 1]
          E <- E[-(k + 1)]
        }
      }
      nbins <- nbins - 1
      if (all(E > 5)) break
    }
    bins
  }
  if(which=="Equal Prob") {kappa <- 0;k <- ifelse(is.null(case$nbins), 10, case$nbins)}
  if(which=="RGd") {kappa <- 0.5;k <- 5+length(case$param)}
  if(which=="Equal Size") {kappa <- 1; k <- ifelse(is.null(case$nbins), 10, case$nbins)}
  case$dta <- x
  if(!is.null(case$param)) case$param <- case$est.mle(x)
  bins <- bin.fun(case, k = k, kappa = kappa)
  tmpbins <- c(-Inf, bins[2:(length(bins)-1)], Inf)
  O <- hist(x, breaks = tmpbins, plot = FALSE)$counts
  E <- length(x)*diff(case$pnull(tmpbins, case$param))
  sum( (O-E)^2/E )
}

\end{lstlisting}

\end{document}